\documentclass[a4paper,fleqn,usenatbib]{mnras}

\usepackage{mathptmx}

\usepackage[T1]{fontenc}
\usepackage{ae,aecompl}

\usepackage{graphicx}	
\usepackage{amsmath}	
\usepackage{amssymb}	
\usepackage{color}
\usepackage{multirow}

\def\apj{ApJ}%
\def\apjl{ApJ}%
\def\apjs{ApJS}%
\def\aap{A\&A}%
\def\mnras{MNRAS}%
\title[Thermal instability]{Thermal instability revisited}

\author[S.A.E.G. Falle et al.]{
S. A. E. G.  Falle$^1$,\thanks{E-mail: sam@amsta.leeds.ac.uk}
C. J. Wareing$^2$
and J. M. Pittard$^2$ 
\\
$^1$Department of Applied Mathematics, University of Leeds, Leeds
LS2 9JT, UK\\
$^{2}$School of  Physics  and  Astronomy,  University of Leeds, Leeds
LS2 9JT, UK
}
\date{Accepted XXX. Received YYY; in original form ZZZ}

\pubyear{2019}

\begin{document}
\label{firstpage}
\pagerange{\pageref{firstpage}--\pageref{lastpage}}
\maketitle

\begin{abstract}
Field's  linear  analysis of  thermal  instability  is repeated  using
methods related to Whitham's theory  of wave hierarchies, which brings
out the physically  relevant parameters in a much clearer  way than in
the  original  analysis.   It  is  also  used  for  the  stability  of
non-equilibrium  states and  we show  that  for gas  cooling behind  a
shock, the usual analysis is only quantitatively valid for shocks that
are just able  to trigger a transition to the  cold phase.  A magnetic
field can  readily be included and  we show that this  does not change
the stability criteria. By considering steady shock solutions, we show
that almost  all plausible initial  conditions lead to  a magnetically
dominated state on the unstable  part of the equilibrium curve.  These
results are used to analyse numerical calculations of perturbed steady
shock solutions and of shocks interacting with a warm cloud.

\end{abstract}


\section{Introduction}
It is very  common for astrophysical plasmas to be  subject to heating
and cooling  processes.  If these  are sufficiently rapid  compared to
other  relevant  timescales,  then  the  plasma  will  be  in  thermal
equilibrium and if this is unstable, then we have a possible mechanism
for  generating   density  inhomogeneities  that  does   not  rely  on
self-gravity.  This motivated \cite{Field:1965} to consider the linear
stability of such  equilibrium states and to apply his  results to the
solar chromosphere  and corona,  planetary nebulae, the  galactic halo
and  galaxy   formation.   As   discussed  in   \cite{Field:1969}  and
\cite{McKee:1977},  thermal instability  is also  a key  ingredient in
multiphase models  of the  ISM.  Since then  there have  been numerous
papers that  have considered  the effect of  thermal instability  in a
diverse  range   of  situations,  such  as   solar  prominences,  e.g.
\cite{Xia:2016},   star   forming   regions,   e.g.   \cite{Kim:2008},
broad-line regions in active  galaxies, e.g.  \cite{Begelman:1990} and
the circumgalactic medium, e.g.   \cite{Stern:2016}.  The analysis has
also been extended to  include perturbations of non-equilibrium states
(e.g. \citealt{Schwarz:1972, Balbus:1986, Koyama:2000}).

Recently,  \cite{Waters:2019}  have  revisited  Field's  analysis  and
written the  dispersion relation in  a somewhat different  form. Their
paper  includes a  discussion  of the  various  modes of  instability,
together with  numerical calculations  of the non-linear  evolution of
the  condensation mode.   They also  consider non-equilibrium  initial
states.

In view of the above, one might  suppose that there is little point in
revisiting  the   linear  analysis   of  either  the   equilibrium  or
non-equilibrium  states. However,  there is  a more  modern method  of
analysing  such  a linear  dispersion  relation  based on  ideas  from
control   theory  and   the   Whitham  theory   of  wave   hierarchies
(\citealt{Whitham:1974}).  Although  this yields  few new  results, it
does illuminate the  physics rather more clearly  than the traditional
approach. 

The analysis  is described in  Section \ref{section_hypbal} and  it is
applied to  the energy source function  proposed by \cite{Koyama:2002}
in  Section \ref{Koyama_en}.   In Section  \ref{section_numcalc} these
results  are   used  to   analyse  numerical  calculations   of  shock
interactions and the work is summarised in Section \ref{section_sum}.

\section{Hyperbolic balance laws}
\label{section_hypbal}

A system of  hyperbolic balance laws in one space  dimension is of the
form

\vspace*{-10pt}
\begin{equation}
\partial_t {\bf u} + \partial_x {\bf f}({\bf u}) = {\bf s}({\bf u}),
\label{hypbal}
\end{equation}
\vspace*{-10pt}

\noindent
where ${\bf  u} =  (u_1, \cdots u_n)^t$   are a  set of $n$  conserved
quantities,   ${\bf f}({\bf  u})   = (f_1,   \cdots  f_n)^t$ are   the
associated fluxes and ${\bf  s}({\bf  u}) =  (s_1, \cdots s_n)^t$  are
source terms  depending upon ${\bf u}$.  Here the superfix $t$ denotes
the transpose.

For sufficently  short wavelengths, the derivatives  dominate over the
source terms  and we have  a frozen system in  which ${\bf s}$  can be
neglected. For long wavelengths the  source term dominates and we have

\vspace*{-10pt}
\begin{equation}
{\bf s} = 0.
\label{equicond}
\end{equation}
\vspace*{-10pt}

This imposes  $r$ conditions  on ${\bf  u}$ where $r$  is the  rank of
${\bf s}$. The system is then reduced to an equilibrium system with $n
- r$ variables  described by  ${\bf u}_e =  {\bf u}_e({\bf  u})$.  The
Whitham theory only considers the case $r  = 1$, but it is possible to
extend it to $r > 1$ (\citealt{Falle:2019}).

\subsection{Gas dynamics with an energy source}

For gas dynamics with an energy source term, we have

\vspace*{-10pt}
\begin{equation}
{\bf u} = [\rho, \rho v, e]^t,~~{\bf f} = [\rho v, p + \rho v^2, v(e +
  p)]^t,~~{\bf s} = [0,0, -\rho L]^t,
\label{gasdyn}
\end{equation}
\vspace*{-10pt}

\noindent
where $\rho$, $v$, $p$ are the density, velocity and pressure. 

\vspace*{-10pt}
\begin{equation}
e = \frac{p}{(\gamma - 1)} + \frac{1}{2} \rho v^2,
\label{toten}
\end{equation}
\vspace*{-5pt}

\noindent
is the  total energy per  unit volume and  $L(\rho, T)$ is  the energy
loss rate per unit mass. Clearly $r = 1$ in this case.

It  is more  convenient to  write these in the form

\vspace*{-10pt}
\begin{equation}
\partial_t {\bf p} + {\rm A} \partial_x {\bf p} = (\gamma - 1) {\bf s},
\label{primeq}
\end{equation}
\vspace*{-10pt}

\noindent
where

\vspace*{-10pt}
\begin{equation}
{\bf p} = [\rho, v, p]^t,
\label{primvar}
\end{equation}
\vspace*{-10pt}

\noindent
are the primitive variables and

\vspace*{-10pt}
\begin{equation}
{\rm A} = \left({
\begin{array}{ccc}
v & \rho & 0\\
0 & v & 1/\rho\\
0 & \gamma p & v\\
\end{array}
}\right).
\label{hypmatrix}
\end{equation}
\vspace*{-5pt}

We now assume a solution of the form

\vspace*{-10pt}
\begin{equation}
{\bf p} = {\bf p}_0 + {\bf p}_1 \exp(i \omega t - i k x),
\label{persol}
\end{equation}
\vspace*{-10pt}

\noindent
where ${\bf  p}_0$ is an  equilibrium state  with $v =  0$, $L(\rho_0,
T_0) =  0$ and  ${\bf p}_1$  is a  small perturbation.  The linearised
equations then give

\vspace*{-10pt}
\begin{equation}
i \omega {\bf p}_1 - i k {\rm A}_0 {\bf p}_1 = {\rm D}_0 {\bf
  p}_1,
\label{perturbeqn}
\end{equation}
\vspace*{-10pt}

\noindent
where $A_0 = A(\rho_0, T_0)$ and

\vspace*{-10pt}
\begin{equation}
{\rm D}_0 = \left({
\begin{array}{ccc}
0 & 0 & 0\\
0 & 0 & 0\\
- \rho_0 G_\rho + p_0 G_T/\rho_0& 0 & -G_T\\
\end{array}
}\right),
\label{dissmatrix}
\end{equation}
\vspace*{-5pt}

\noindent
with

\vspace*{-10pt}
\begin{equation}
\begin{array}{rclrcl}
G_\rho   &  =   &   (\gamma  -   1)   L_\rho  &   =   &(\gamma  -   1)
\displaystyle{\left( \frac{\partial L}{\partial \rho} \right)_T},\\
& & \\
G_T  &  =  & (\gamma  -  1)  L_T  &  = &  \displaystyle{(\gamma  -  1)
  \frac{m}{k_B}\left( \frac{\partial L}{\partial T} \right)_\rho},\\
\end{array}
\label{Gdef}
\end{equation}
\vspace*{-5pt}

\noindent
where $k_B$ is Boltzmann's constant and $m$ is the mean particle mass.
The eigenvalues of ${\rm D}$ are $0$,  $0$ and $-G_T$, so that we have
isochoric instability if

\vspace*{-10pt}
\begin{equation}
G_T < 0 \Rightarrow L_T < 0,
\label{isochoric}
\end{equation}
\vspace*{-10pt}

\noindent
which  is equation  (4a) in  \cite{Field:1965}.  From now  on we  will
assume that $G_T > 0$.

The dispersion relation associated with (\ref{perturbeqn}) is

\vspace*{-10pt}
\begin{equation}
|\omega {\rm I} - k {\rm A}_0 + i {\rm D}_0| = 0,
\label{disprel}
\end{equation}
\vspace*{-10pt}

\noindent
where ${\rm I}$ is the identity matrix. This can be written

\vspace*{-10pt}
\begin{equation}
P = P_0 - \frac{i}{k} P_1 = 0, 
\label{dispoly}
\end{equation}
\vspace*{-10pt}

\noindent
where

\vspace*{-10pt}
\begin{equation}
P_0  = \mu  (\mu^2 -  a_f^2),~~P_1 =  G_T(\mu^2 -  a_e^2), ~~  \mu =
  \frac{\omega}{k}.
\label{polys}
\end{equation}
\vspace*{-10pt}

\noindent
Here the frozen sound speed, $a_f$,  applies in the adiabatic case and
the  equilibrium sound  speed, $a_e$,  when the  system is  in thermal
equilibrium i.e.   equation (\ref{equicond})  is satisfied.  These are
given by

\vspace*{-10pt}
\begin{equation}
a_f^2 =  \frac{\gamma p_0}{\rho_0},  ~~ a_e^2 =  \left( \frac{\partial
  p_0}{\partial  \rho_0  }  \right)_{L=0} =  \frac{G_T  a_f^2  -\rho_0
  \gamma G_\rho}{\gamma G_T}.
\label{soundspeeds}
\end{equation}
\vspace*{-5pt}

$P_0$ describes the frozen  system since its roots are the
wave  speeds of  the  frozen  system, $0$,  $\pm  a_f$, whereas  $P_1$
describes the equilibrium  system since its roots, $\pm  a_e$, are the
wave speeds of the equilibrium system.  If we define the
acoustic or thermal wavenumber by

\vspace*{-10pt}
\begin{equation}
k_T = \frac{G_T}{a_f} = \frac{(\gamma - 1) m L_T}{k_B a_f},
\label{thermk}
\end{equation}
\vspace*{-5pt}

\noindent
then the  source term  is neglible  for $k  \gg k_T$  and we  have the
frozen  system   whereas  the  source  term   dominates  and  enforces
equilibrium for  $k \ll k_T$.  This  is the same as  $k_T$ in equation
(16) in \cite{Field:1965}.  The corresponding wavelength is

\vspace*{-10pt}
\begin{equation}
\lambda_T =  \frac{2 \pi}{k_T} = \frac{2  \pi k_B a_f}{(\gamma -  1) m
  L_T}.
\label{thermlambda}
\end{equation}
\vspace*{-5pt}

This   separation  of   the  dispersion   relation  into   polynomials
corresponding to the frozen and  equilibrium systems was first used by
\cite{Whitham:1974} in his theory of wave hierarchies. As he shows, it
can be  applied to many  different systems,  but to our  knowledge has
only been used in an astrophysical context by \cite{Tytarenko:2002}.

The   Hermite-Biehler   theorem  (e.g.    \citealt{Bhattacharyya:1995,
  Tytarenko:2002, Falle:2019})  tells us  that if the  coefficients of
the highest power of $\mu$ in $P_0$  and $P_1$ have the same sign, the
roots of  $P_0$ and $P_1$ are  real and the roots  of $P_1$ interleave
with  those of  $P_0$,  then  the roots  of  (\ref{dispoly}) all  have
positive imaginary  parts and  the system  is stable.   This stability
condition is both necessary and sufficient.

The system will  certainly be unstable if $a_e^2$  is imaginary, which
requires

\vspace*{-10pt}
\begin{equation}
a_e^2  =  \frac{G_T  a_f^2  -\rho_0   \gamma  G_\rho}{\gamma  G_T}  <  0
~\Rightarrow~ 1 - \frac{\rho_0}{T_0} \frac{L_{\rho}}{L_T} < 0,
\label{isobaric}
\end{equation}
\vspace*{-5pt}

\noindent
from equation  (\ref{soundspeeds}). This  is the  isobaric instability
condition, equation (4b) in \cite{Field:1965} when $L_T > 0$.

For real $a_e$, the roots of $P_0$ and $P_1$ do not interleave if

\vspace*{-10pt}
\begin{equation}
a_e > a_f \Rightarrow a_f^2 - a_e^2 =  (\gamma - 1) \frac{p_0}{\rho_0} +
\frac{\rho_0 G_\rho}{G_T} = (\gamma - 1) \frac{p_0}{\rho_0} +
\frac{k_B \rho_0 L_\rho}{m L_T} < 0.
\label{iseninstab}
\end{equation}
\vspace*{-10pt}

\noindent
This is the isentropic instability condition for sound waves, equation
(5) in \cite{Field:1965} when $L_T > 0$.

In the absence of conduction, the only dimensionless parameter is

\vspace*{-10pt}
\begin{equation}
\alpha =  \frac{a_e^2}{a_f^2}.
\label{alpha}
\end{equation}
\vspace*{-5pt}

\noindent
This plays the same role  as the dimensionless parameter, $\alpha$, in
\cite{Field:1965}, but  has a  more obvious physical  significance. In
particular, the stability conditions take the simple form

\vspace*{-10pt}
\begin{equation}
\begin{array}{ll}
\alpha < 0 & \mbox{isobaric~instability,} \\    
0 \le \alpha \le 1 & \mbox{stable,}\\
\alpha > 1 & \mbox{isentropic~instability.}
\end{array}
\label{stabcon}
\end{equation}
\vspace*{-5pt}

Now consider the behaviour of the root,  $\mu = -a_e$, for small $k$. We
get

\vspace*{-10pt}
\begin{equation}
\omega = -a_e  k + i k^2  \frac{(a_f^2 - a_e^2)}{2 G_T}  + k^3 \frac{5
  a_e^4 - 6 a_e^2 a_f^2 + a_f^4}{8 G_T^2 a_f} + O(k^4),
\label{smallk}
\end{equation}
\vspace*{-5pt}

\noindent
which corresponds to instability when $a_e^2 < 0$.

Similarly, for large $k$, the root $\mu = 0$, which corresponds to the
non-propagating condensation mode, is given by

\vspace*{-10pt}
\begin{equation}
\omega    =    \frac{i    a_e^2   G_T}{a_f^2}    +    \frac{i    a_e^4
  G_T^3(a_f^2-a_e^2)}{k^2 a_f^8} + O(1/k^4).
\label{largek}
\end{equation}
\vspace*{-5pt}

\noindent
Again we have  instability if $a_e^2 < 0$. The  first term agrees with
equation (31) in \cite{Field:1965}. This tells us that the growth rate
tends to  a constant as  $k \rightarrow  \infty$ and that  the largest
wavenumber modes are the most unstable.

\subsection{Thermal conduction}

If  thermal conduction  is important,  then the  pressure equation  in
(\ref{primeq}) becomes

\vspace*{-10pt}
\begin{equation}
\partial_t p + \gamma  p \partial_x v + v \partial_x p  = (\gamma - 1)
\partial_x \kappa \partial_x T
\end{equation}
\vspace*{-10pt}

\noindent
where  $\kappa$  is  the thermal  conductivity.  (\ref{disprel})  then
becomes

\vspace*{-10pt}
\begin{equation}
|\omega {\rm I} - k {\rm A}_0 + i {\rm D}_0 - i k^2 {\rm C}_0| = 0,
\label{conddisprel}
\end{equation}
\vspace*{-10pt}

\noindent
where

\vspace*{-10pt}
\begin{equation}
{\rm C}_0 = \left({
\begin{array}{ccc}
0 & 0 & 0\\
0 & 0 & 0\\
- \kappa^\prime p_0/ \rho_0^2 & 0 & \kappa^\prime/\rho_0\\
\end{array}
}\right),
\label{condmatrix}
\end{equation}
\vspace*{-5pt}

\noindent
with

\vspace*{-10pt}
\begin{equation}
\kappa^\prime = (\gamma - 1) \frac{m}{k_B} \kappa.
\label{kappaprime}
\end{equation}
\vspace*{-10pt}

\noindent
Equation (\ref{dispoly}) becomes

\vspace*{-10pt}
\begin{equation}
P =  P_0 - \frac{i}{k}  P_1 - i k  P_2 = 0,
\label{conddisp}
\end{equation}
\vspace*{-5pt}

\noindent
where

\vspace*{-10pt}
\begin{equation}
P_2 = \frac{\kappa^\prime}{\rho_0}  (\mu^2 - a_f^2/\gamma) =
\frac{\kappa^\prime}{\rho_0}  (\mu^2 - c_T^2),
\label{condpoly}
\end{equation}
\vspace*{-10pt}

\noindent
and

\vspace*{-10pt}
\begin{equation}
c_T = \left( \frac{p_0}{\rho_0} \right)^{1/2} 
\label{isosound}
\end{equation}
\vspace*{-5pt}

\noindent
is the  isothermal sound speed.  One  would expect $P_2$ to  have this
form since  it describes the  behaviour when  $k$ is large  enough for
thermal conduction to ensure a  uniform temperature. We now have three
polynomials each  associated with a different  physical process: $P_0$
for the adiabatic system, $P_1$  when the energy source term dominates
and $P_2$  when thermal conduction  dominates.  \cite{Liubarskii:1961}
calls these the auxiliary polynomials.

For $\gamma  > 1$, the roots  of $P_2$ interleave with  those of $P_0$
and thermal conduction is stabilising. If the system is subject to the
isobaric  instability,  then $a_e^2  <  0$  and conduction  stabilises
wavenumbers for which

\[
\frac{G_T a_e^2}{k} + \frac{k \kappa^\prime a_f^2}{\rho_0 \gamma} > 0,
\]

\noindent
i.e.

\vspace*{-10pt}
\begin{equation}
k > k_F = \left(  \frac{-a_e^2 \gamma \rho_0 G_T}{a_f^2 \kappa^\prime}
\right)^{1/2}  =  \left(  \frac{-\alpha \gamma  \rho_0  L_T  }{\kappa}
\right)^{1/2}.
\label{fieldk1}
\end{equation}
\vspace*{-5pt}

\noindent
The corresponding Field length is then

\vspace*{-10pt}
\begin{equation}
\lambda_F  = \frac{2  \pi}{k_F} =  2 \pi  \left( \frac{\kappa}{-\alpha
  \gamma \rho_0 L_T} \right)^{1/2},
\label{fieldlambda}
\end{equation}
\vspace*{-5pt}

\noindent
which  agrees   with  equation  (26)  in   \cite{Field:1965}  and  the
expressions     in    \cite{Begelman:1990}     and    \cite{Kim:2008}.
On the other hand, \cite{Koyama:2004} define

\vspace*{-10pt}
\begin{equation}
\lambda_F = \left( \frac{T_0 \kappa}{\rho_0 L_c} \right)^{1/2},
\label{koyamalambda}
\end{equation}
\vspace*{-5pt}

\noindent
where  $L_c$  is the  magnitude  of  the  cooling  term in  $L$.   The
advantage  of (\ref{fieldlambda})  is  that it  really  is the  linear
stability   limit.  (\ref{fieldlambda})  and (\ref{koyamalambda}) are
in   fact   very   different   since  the   $\lambda_F$   defined   by
(\ref{fieldlambda}) goes to infinity at the boundaries of the unstable
region, ($a_e = 0$), as it should.

The effect of conduction is determined by the dimensionless parameter

\vspace*{-10pt}
\begin{equation}
\beta = \frac{G_T \kappa^\prime}{\rho a_f^2} = \frac{(\gamma -1)^2 m^2
  L_T \kappa}{k_B^2 \rho_0 a_f^2}.
\label{beta}
\end{equation}
\vspace*{-5pt}

\noindent
The Field wavenumber is then given by

\vspace*{-10pt}
\begin{equation}
k_F = k_T \left( \frac{- \gamma \alpha}{\beta} \right)^{1/2}.
\label{fieldk2}
\end{equation}
\vspace*{-5pt}

\noindent
Again, this seems  to be a more natural choice  than the corresponding
dimensionless parameter, $\beta$, in \cite{Field:1965}.

If we define the dimensionless variables

\vspace*{-10pt}
\begin{equation}
\bar{\mu}  =   \frac{\mu}{a_f},~~  \bar{k}   =  \frac{k   a_f}{G_T}  =
\frac{k}{k_T},
\label{dimvar}
\end{equation}
\vspace*{-5pt}

\noindent
then (\ref{conddisp}) becomes

\vspace*{-10pt}
\begin{equation}
\bar{\mu} (\bar{\mu}^2 - 1) - \frac{i}{\bar{k}} (\bar{\mu}^2 -
\alpha) - i \beta \bar{k} (\bar{\mu}^2 - 1/\gamma) = 0.
\label{dimpoly}
\end{equation}
\vspace*{-5pt}

\noindent
If we put

\vspace*{-10pt}
\begin{equation}
\bar{\mu} = - i y,
\label{ydef}
\end{equation}
\vspace*{-5pt}

\noindent
then (\ref{dimpoly}) becomes

\vspace*{-10pt}
\begin{equation}
y (y^2 +  1) + \frac{1}{\bar{k}} (y^2 + \alpha)  + \beta \bar{k}
(y^2 + 1/\gamma) = 0,
\label{Fieldpoly}
\end{equation}
\vspace*{-5pt}

\noindent
which is our version of  equation (18) in \cite{Field:1965}. Note that
equation  (\ref{fieldk2}) tells  us that  this  has a  zero root  when
$\bar{k} = k_F / k_T$, as expected.

\subsection{Magnetic field}

The analysis  can readily be  extended to include an  oblique magnetic
field  with  components $B_x$,  $B_y$.   We  have  the fast  and  slow
magnetosonic speeds,

\begin{equation}
c_{f,s}^2 = \frac{1}{2} \left[ {a^2 + B^2 / \rho \pm \surd
    \left\{ {(a^2 + B^2 / \rho)^2 - 4 B_x a^2 / \rho } \right\} }
  \right],
\label{magnetosonic}
\end{equation}

\noindent
where  $a  =  a_f$ for  the  frozen  system  and  $a =  a_e$  for  the
equilibrium system.

In the absence of conduction, the dispersion relation must now be

\vspace*{-10pt}
\begin{equation}
\mu (\mu^2 - c_{ff}^2) (\mu^2 - c_{fs}^2) - \frac{i}{k} G_T (\mu^2 -
c_{ef}^2) (\mu^2 - c_{es}^2) = 0, 
\label{mdisp}
\end{equation}
\vspace*{-5pt}

\noindent
where $c_{ff}$, $c_{fs}$ are the frozen fast/slow speeds and $c_{ef}$,
$c_{es}$ the  equilibrium ones.  Since  the equilibrium slow  speed is
imaginary when $a_e$  is imaginary, the isobaric  instability is still
given by  (\ref{stabcon}).  Furthermore,  the interleaving  also fails
when $a_e >  a_f$, so that isentropic instability is  also governed by
(\ref{stabcon}) i.e.   the stability  conditions are  unchanged.  More
surprisingly, the growth rate for large $k$ is now

\vspace*{-10pt}
\begin{equation}
\omega = \frac{i G_T  c_{ef}^2 c_{es}^2}{c_{ff}^2 c_{fs}^2} + O(1/k^2)
= \frac{i a_e^2 G_T }{a_f^2} + O(1/k^2)
\label{mlargek}
\end{equation}
\vspace*{-5pt}

\noindent
i.e.  exactly the same as (\ref{largek}) for the non-magnetic case. It
agrees with the result in \cite{Dudorov:2019}.  Note that for a purely
transverse field  the slow speed  is zero and  it is possible  for the
magnetic  field to  stabilise the  isobaric mode.  However, this  is a
singular case that has a vanishingly small probability of occurring in
reality.

This is another  illustration of the power of the  method: the physics
tells how to write down the adiabatic and equilibrium polynomials from
what  we  already  know  about  the wave  speeds  of  the  frozen  and
equilibrium MHD  systems.  As  we have already  pointed out,  we could
also have done this for thermal conduction.  The only difficulty is in
obtaining the  coefficient multiplying the polynomials,  but these can
often  be obtained  by inspection  of  the relevant  matrix.  This  is
certainly true for the energy  source, thermal conduction and magnetic
field.

\subsection{Stability of non-equilibrium states}
\label{Nonequistab}

\cite{Field:1965},    \cite{Schwarz:1972},   \cite{Balbus:1986}    and
\cite{Koyama:2000} extend the thermal instability analysis to gas that
is  not in  thermal equilibrium.  \cite{Schwarz:1972} assume  that the
unperturbed   density  is   constant,   \cite{Koyama:2000}  that   the
unperturbed  pressure is  constant and  \cite{Balbus:1986} consider  a
general unperturbed state.  \cite{Balbus:1986} and \cite{Schwarz:1972}
go somewhat  further than \cite{Koyama:2000}  in that they use  a JWKB
approximation to take account of the time variation of the unperturbed
state.  However, in  all cases the analysis is local  i.e.  only valid
in the short wavelength limit.

\cite{Koyama:2000}  assume that  the gas  is contracting  uniformly so
that lengths scale like $R(t)$. They introduce a scaled coordinate

\vspace*{-10pt}
\begin{equation}
  \bar{x} = \frac{x}{R},
\label{xbar}
\end{equation}
\vspace*{-10pt}

\noindent
with $R(0) = 1$. The primitive equations, (\ref{primeq}),  become

\vspace*{-10pt}
\begin{equation}
\partial_t  {\bf  p}  +  \frac{1}{R}  A  \partial_{\bar{x}} {\bf  p}  =
\frac{1}{R^2} \partial_{\bar{x}} {\bf c} + {\bf s} ,
\label{isoprimeqn}
\end{equation}
\vspace*{-5pt}

\noindent
with

\vspace*{-10pt}
\begin{equation}
{\bf c} = [0,0, (\gamma - 1) \kappa \partial_{\bar{x}} T]^t,
\label{isocond}
\end{equation}
\vspace*{-5pt}

\noindent
and

\vspace*{-10pt}
\begin{equation}
{\bf   s}   =   [-\frac{\dot{R}}{R}    \rho,   -\ddot{R}   \bar{x}   -
  \frac{\dot{R}}{R} v, - \frac{\dot{R}}{R} \gamma p -(\gamma - 1) \rho
  L]^t.
\label{isosource}
\end{equation}
\vspace*{-5pt}

\noindent
The velocity,  $v$, in ${\bf  s}$ and $A$ is  now the velocity  in the
co-moving frame $v \rightarrow v - \dot{R} \bar{x}$.

They  then  consider  a  spatially uniform  unperturbed  state,  ${\bf
  p}_0(t)$, with  constant pressure, $p_0$,  and zero velocity  in the
contracting frame. This satisfies

\vspace*{-10pt}
\begin{equation}
p_0(t)   =   p_0(0),~~\rho_0(t)  =   \frac{\rho_0(0)}{R(t)},~~v(t)   =
0,~~\frac{\dot{R}}{R}  = -  \frac{(\gamma  - 1)}{\gamma}  \frac{\rho_0
  L}{p_0}.
\label{isounpert}
\end{equation}
\vspace*{-10pt}

\noindent
Integrating the  last of  these equations gives  $R(t)$ and  hence the
solution. Note  that this is only  valid for regions small  enough for
the  term $\ddot{R}  \bar{x}$  to be  negligible,  which requires  short
wavelengths.

We assume a perturbation of the form

\vspace*{-10pt}
\begin{eqnarray}
\rho(x,  t) &  =  &  \rho_0(t) [1  +  \rho_1 \exp(i  \omega  t  - i  k
  \bar{x})],\nonumber\\
p(x,t) & = & p_0 [1 + p_1 \exp (i \omega t - i k \bar{x}) ],\\
v(x,t) & = & v_1 \exp (i \omega t  -  i   k \bar{x}),\nonumber
\label{isopersol}
\end{eqnarray}
\vspace*{-10pt}

\noindent
where $\rho_1$, $p_1$ and $v_1$  are constants, which is equivalent to
that used by \cite{Koyama:2000}. Putting this into (\ref{isoprimeqn}),
linearising and neglecting $\ddot{R} \bar{x}$ gives

\vspace*{-10pt}
\begin{equation}
i \omega  {\bf p}_1 -  i \frac{k}{R} {\rm A}_c  {\bf p}_1 =  {\rm D}_c
{\bf p}_1 -\frac{k^2}{R^2} {\rm C}_c {\bf p}_1,
\label{isopereqn}
\end{equation}
\vspace*{-5pt}  

\noindent
where ${\bf p}_1 = (\rho_1, v_1, p_1)^t$,

\vspace*{-10pt}
\begin{equation}
{\rm A}_c = \left({
\begin{array}{ccc}
0 & 1 & 0\\
0 & 0 & p_0/\rho_0\\
0 & \gamma & 0\\
\end{array}
}\right),
\label{isoAmatrix}
\end{equation}
\vspace*{-5pt}

\vspace*{-10pt}
\begin{equation}
{\rm D}_c = \left({
\begin{array}{ccc}
0 & 0 & 0\\
0 & - \sigma_c & 0\\
(-\rho_0 G - \rho_0^2 G_{\rho} + p_0  G_T)/p_0 & 0 & - \gamma \sigma_c
-G_T\\
\end{array}
}\right),
\label{isoDmatrix}
\end{equation}
\vspace*{-5pt}

\noindent
and

\begin{equation}
{\rm C}_c = \left({
\begin{array}{ccc}
0 & 0 & 0\\
0 & 0 & 0\\
- \kappa^\prime / \rho_0 & 0 & \kappa^\prime / \rho_0\\
\end{array}
}\right).
\label{isoCmatrix}
\end{equation}
\vspace*{-5pt}

\noindent
Here

\vspace*{-10pt}
\begin{equation}
G = (\gamma - 1) L
\label{isoGdef}
\end{equation}
\vspace*{-10pt}

\noindent
and

\vspace*{-10pt}
\begin{equation}
\sigma_c = \frac{\dot{R}}{R}.
\label{isosigma_c}
\end{equation}
\vspace*{-5pt}

We can set $R = 1$ since we are only interested in the stability of
the original state. The dispersion relation is then

\vspace*{-10pt}
\begin{equation}
|\omega {\rm  I} - k {\rm  A}_c + i  {\rm D}_c - k^2  i {\rm
  C}_c| = 0,
\label{isodisprel}
\end{equation}
\vspace*{-5pt}

\noindent
which we can write as 

\vspace*{-10pt}
\begin{equation}
P = P_r -  i P_i. 
\label{isodispoly}
\end{equation}

\noindent
In the previous  subsections we showed that it is  useful to split the
dispersion  relation into  polynomials associated  with the  different
physical  processes,  the  adiabatic  system, the  energy  source  and
thermal  conduction.    There  are   now  four   different  processes:
adiabatic, energy source, thermal conduction and the source due to the
isobaric contraction.

We therefore write equation (\ref{isodispoly}) as

\vspace*{-10pt}
\begin{equation}
P_r = P_0 - \frac{1}{k^2} (P_{13} +  P_{33}) - P_{23},~~~~P_i = 
\frac{1}{k} (P_1 +P_3) +  kP_2,
\label{isopolys}
\end{equation}

\noindent
with   $P_0$,  and   $P_2$  given   by  equations   (\ref{polys})  and
(\ref{condpoly}) and

\vspace*{-10pt}
\begin{equation}
\begin{array}{l}
\displaystyle{P_1  = G_T  (\mu^2  -  a_e^2) +  G,~~~  P_3 =  \sigma_c
(\gamma + 1) \mu^2,}\\
~ \\
\displaystyle{P_{13}  = \sigma_c  G_T  \mu,~~~P_{23} =  \frac{\sigma_c
    \kappa}{\rho_0} \mu,~~~P_{33} = \gamma \sigma_c^2 \mu.} \\
\end{array}
\label{isoauxpolys}
\end{equation}

Here the suffices $1$, $2$, $3$ are associated with the energy source,
thermal  conduction and  the isobaric  contraction respectively.   The
dispersion relation is split  into auxiliary polynomials $P_1$, $P_2$,
$P_3$  due to  each process  in isolation,  $P_{13}$, $P_{23}$  due to
interactions between them and $P_{33}$  due to self-interaction of the
isobaric  contraction.   Note  that  the isobaric  contraction  has  a
self-interaction  because it  affects both  the velocity  and pressure
equation,  whereas the  other processes  only appear  in the  pressure
equation.

If we ignore  conduction, then for the condensation mode  at large $k$
we get

\vspace*{-10pt}
\begin{equation}
\omega = i G_T\frac{(a_e^2 - G/G_T)}{a_f^2} + O(1/k^2).
\label{isolargek}
\end{equation}
\vspace*{-5pt}

\noindent
This clearly  also applies to  the magnetic case  if the field  is not
exactly perpendicular.  It is a reasonable approximation to the growth
rate of  the most  unstable short wavelength  mode whenever  the Field
length  is significantly  smaller  than the  acoustic wavelength.   We
therefore have short wavelength instability when

\vspace*{-10pt}
\begin{equation}
a_e^2 G_T - G < 0,
\label{isoinstabcon}
\end{equation}
\vspace*{-5pt}

\noindent
which is  just the isobaric instability  condition for non-equilibrium
states given by \cite{Balbus:1986}.

In order to  determine when (\ref{isolargek}) is  a good approximation
to  the  maximum growth  rate,  we  need  the  Field length  for  this
case. The coefficient of $\mu^2$ in $P_i$ is

\vspace*{-10pt}
\begin{equation}
\frac{1}{k} \left[ { G_T + \sigma_c (\gamma + 1)} \right] +
\frac{k \kappa^\prime}{\rho _0},
\label{mu2coeff}
\end{equation}
\vspace*{-5pt}

\noindent
and this must be positive for stability. $P_i$ has real roots if

\vspace*{-10pt}
\begin{equation}
k^2 > \frac{\rho_0}{\kappa^\prime c_T^2} (G - G_T a_e^2) G_T.
\label{P_irootcon}
\end{equation}
\vspace*{-5pt}

\noindent
In combination, (\ref{mu2coeff}) and (\ref{P_irootcon}) tell us that
we have stability if

\vspace*{-10pt}
\begin{equation}
k > k_f = { \left[  \frac{\rho_0}{\kappa^\prime} {max \left\{ {- G_T -
        \sigma_c  (\gamma  + 1),  \frac{(G  -  G_T a_e^2)  }{c_T^2}  }
      \right\} } \right] }^{1/2}
\label{kfieldne}
\end{equation}
\vspace*{-5pt}

\noindent
This replaces the expression (\ref{fieldk1}) for the Field wavenumber.
It  is everywhere  much larger  than  the thermal  wavenumber for  any
plausible form of $L$, such as the one considered in the next section.

Although  \cite{Koyama:2000} only  considered an  isobaric unperturbed
state,  their  analysis  is  valid   for  short  wavelengths  and  any
unperturbed  state, provided  the  growth rate  is  large compared  to
$|\sigma_c| =  |\dot{R}/R|$.  In  particular, for the  isochoric state
considered  by \cite{Schwarz:1972},  the equations  are the  same with
$\sigma_c = 0$,  $R = 1$, $p_0  = p_0(t)$ and $\rho_0  = {\rm const}$.
Since (\ref{isolargek})  is independent  of $\sigma_c$,  the condition
(\ref{isoinstabcon}) also applies in this case.

\begin{figure}
\includegraphics*[viewport=0 30 375 350,width=1.0\columnwidth]{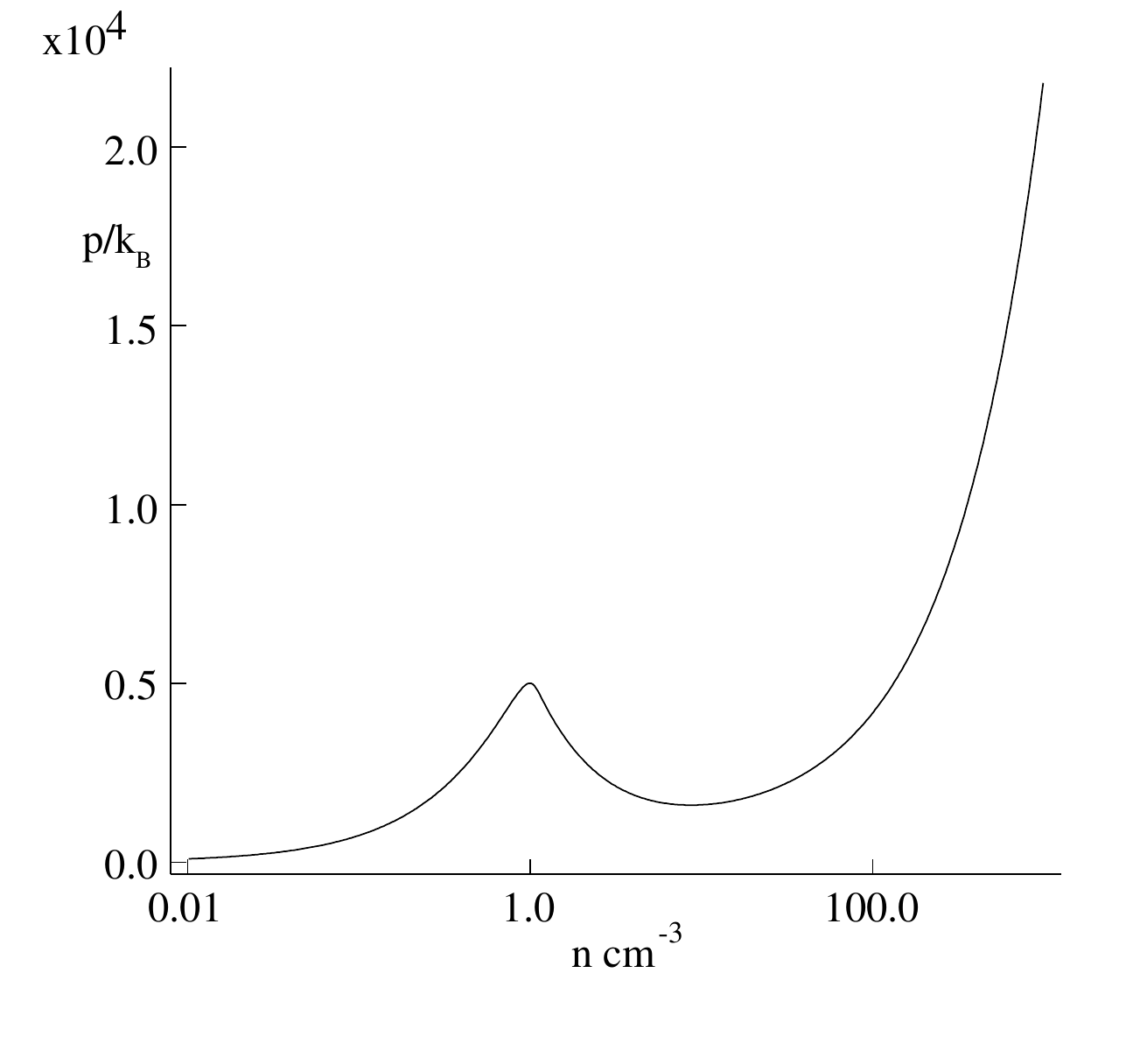}
\caption{The pressure divided by the Boltzmann constant, $k_B$, on the
  equilibrium curve as a function of particle density.}
\label{fig1}
\end{figure}

\begin{figure}
\includegraphics*[viewport=0 30 375 350,width=1.0\columnwidth]{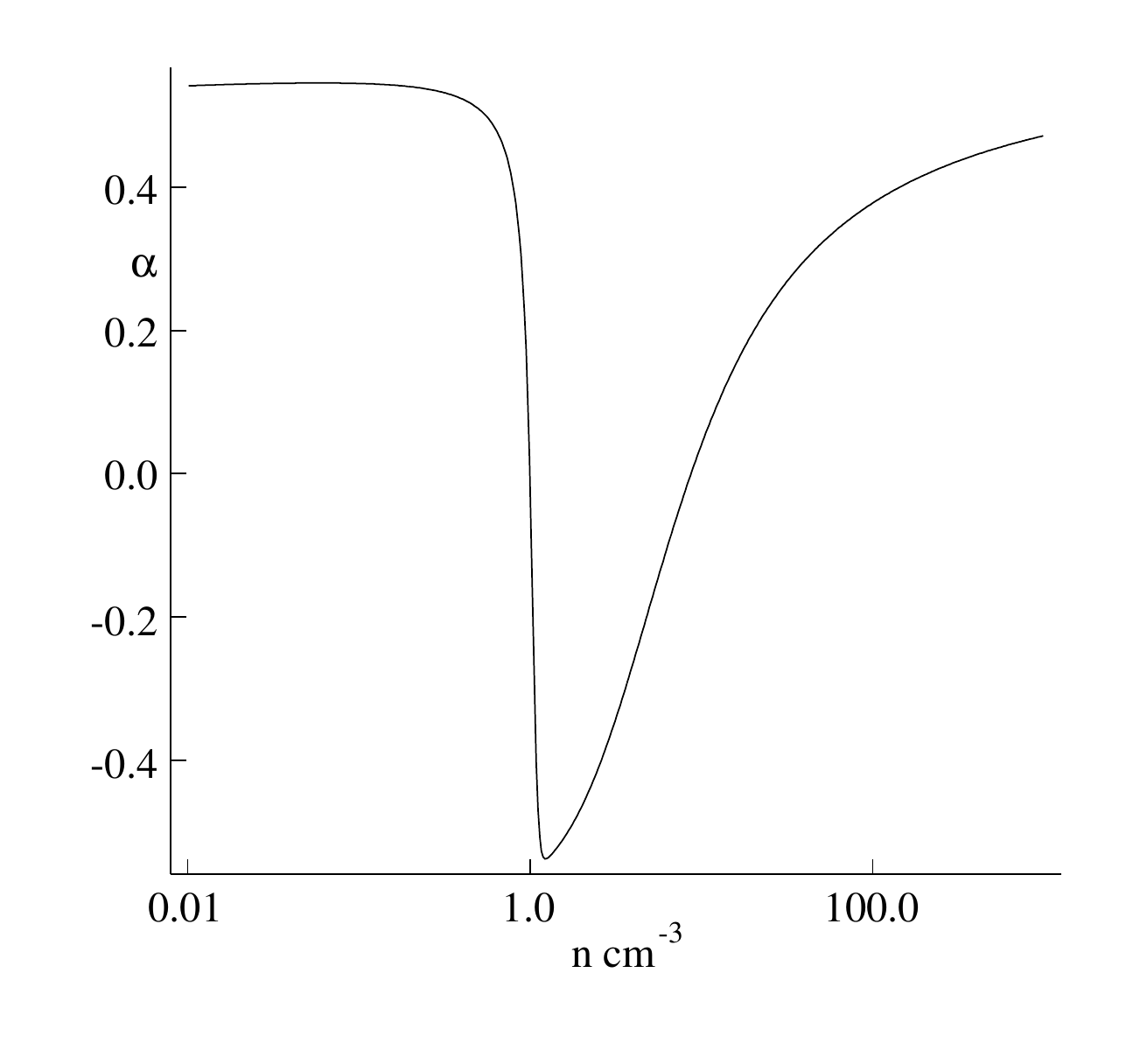}
\caption{The  dimensionless parameter,  $\alpha$, defined  by equation
  (\ref{alpha}) as a function of particle density.}
\label{fig2}
\end{figure}

\begin{figure}
\includegraphics*[viewport=0 30 375 350,width=1.0\columnwidth]{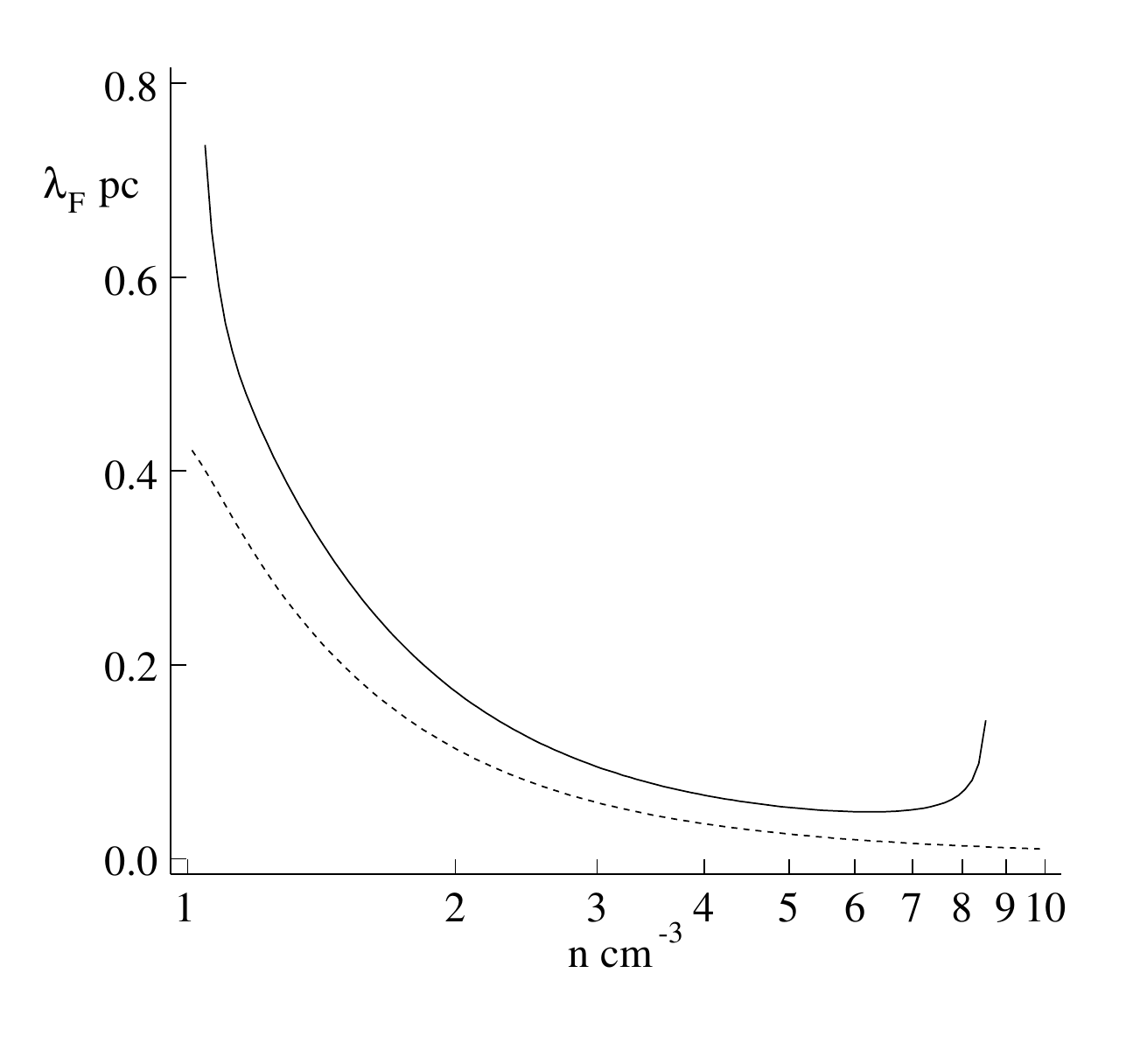}
\caption{The  Field length  as  a function  of  particle density  from
  (\ref{fieldlambda})  (solid  line)   and  from  (\ref{koyamalambda})
  (dashed line).}
\label{fig3}
\end{figure}

\begin{figure}
\includegraphics*[viewport=0 30 375 350,width=1.0\columnwidth]{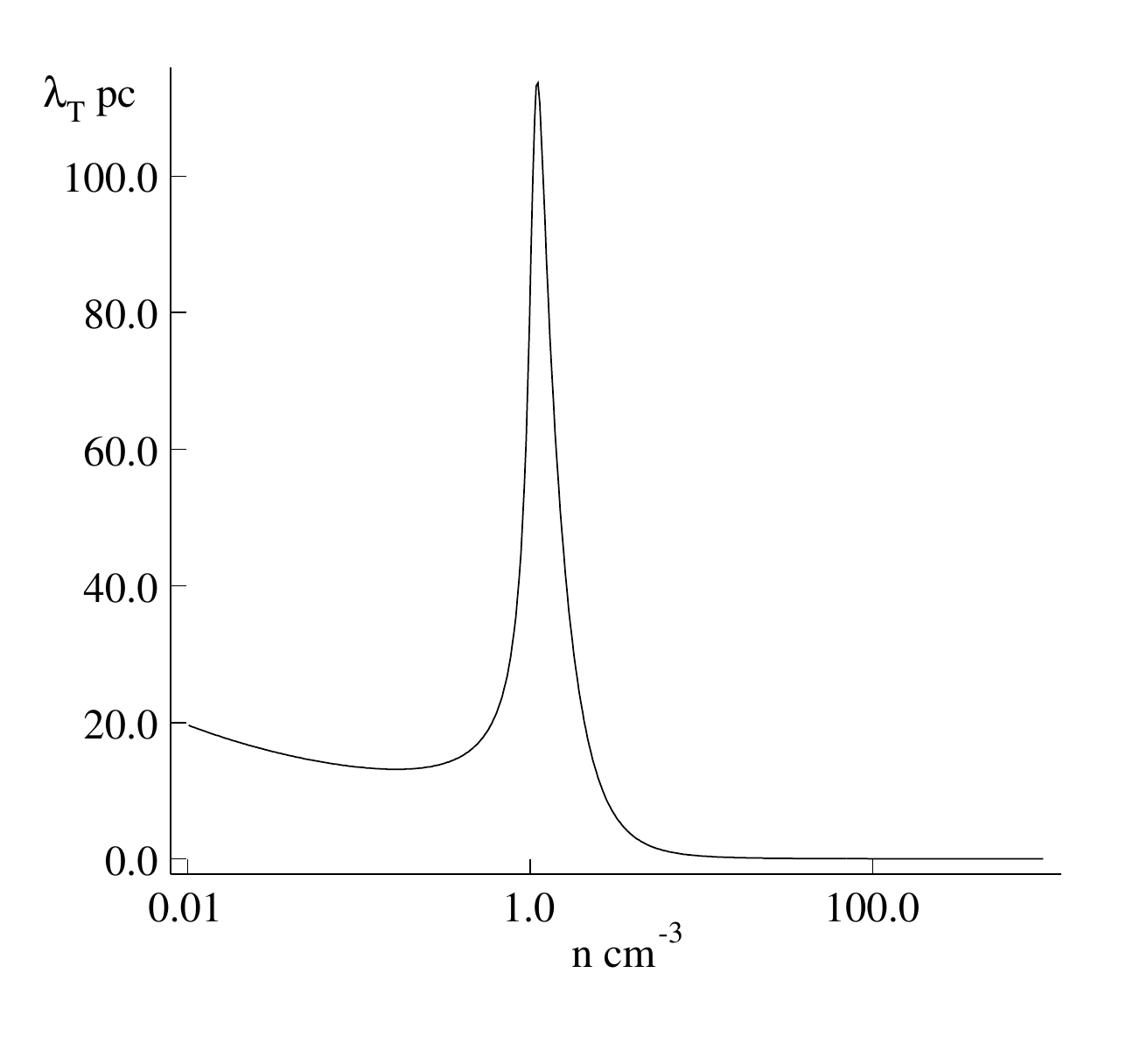}
\caption{The thermal  length as  a function  of particle  density from
  (\ref{thermlambda}).}
\label{fig4}
\end{figure}

\begin{figure}
\begin{tabular}{l}
(a)\\
  \includegraphics[width=8cm]{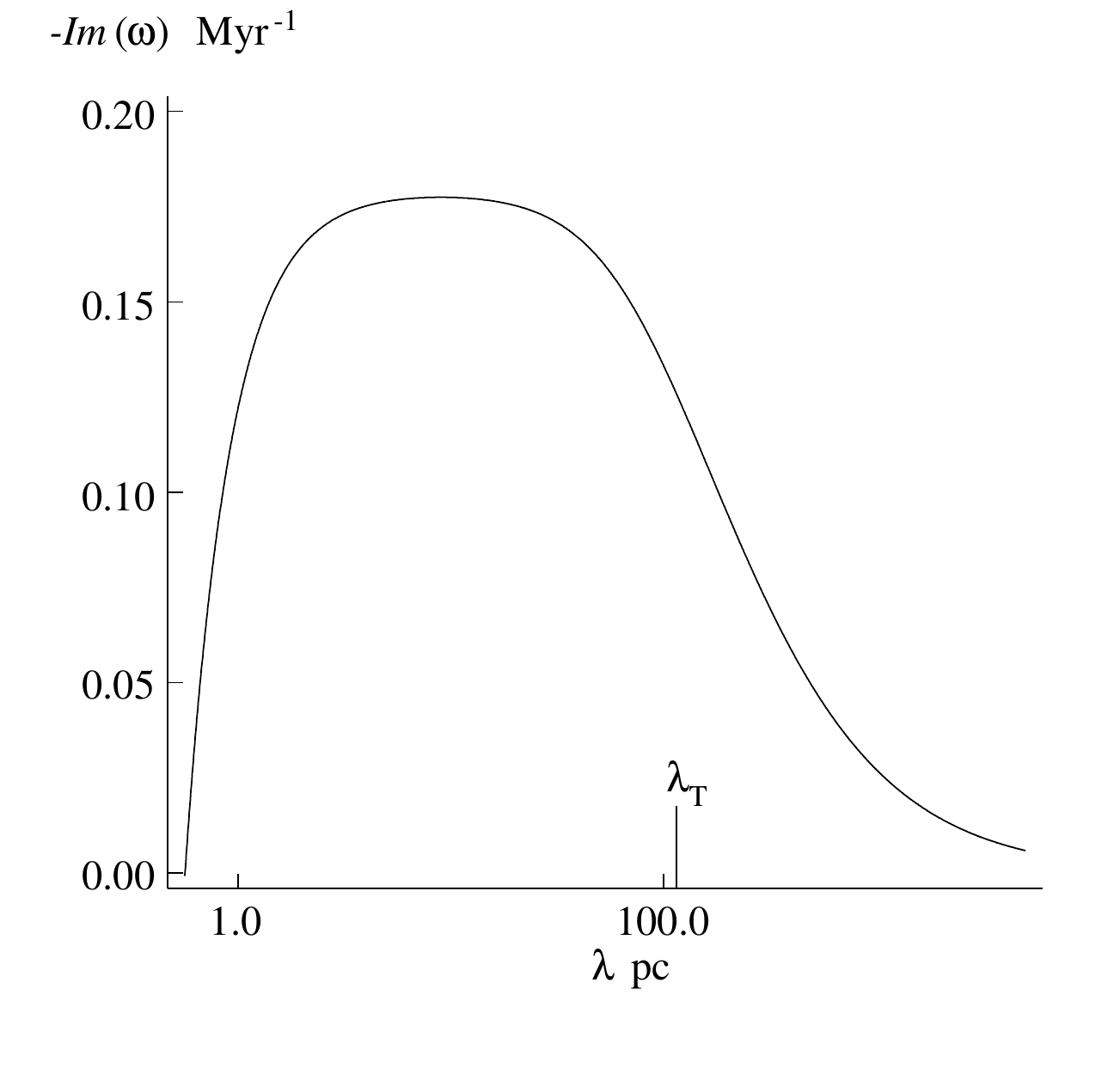} \\
(b)\\
  \includegraphics[width=8cm]{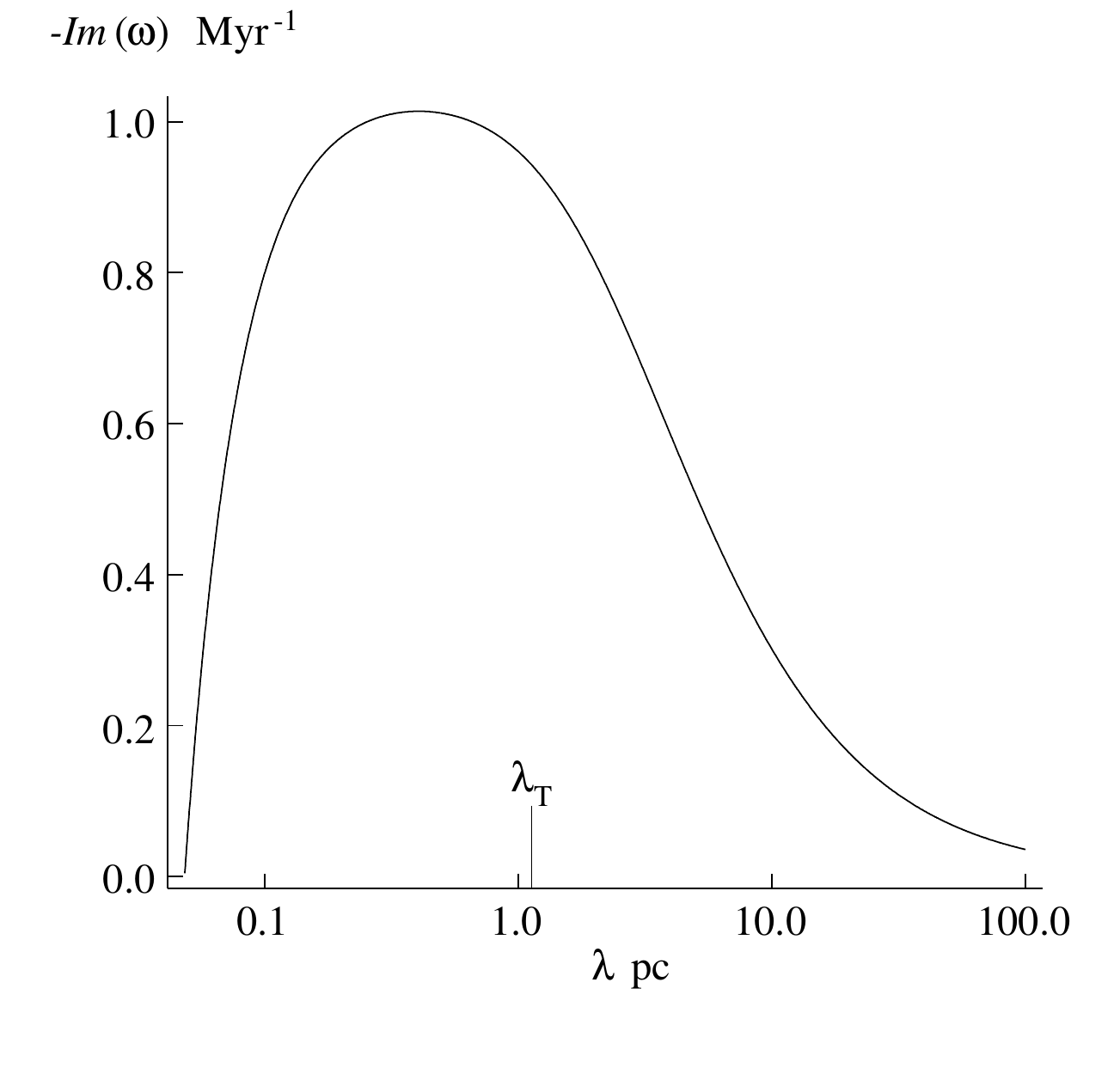}    
\end{tabular}
\caption{The growth  rate of  the condensation mode  as a  function of
  wavelength. a) $n = 1.1$, b) $n  = 6.2517$ (the density at which the
  Field length is a minimum).}
\label{fig5}
\end{figure}

\begin{figure}
\includegraphics*[viewport=0 30 375 350,width=1.0\columnwidth]{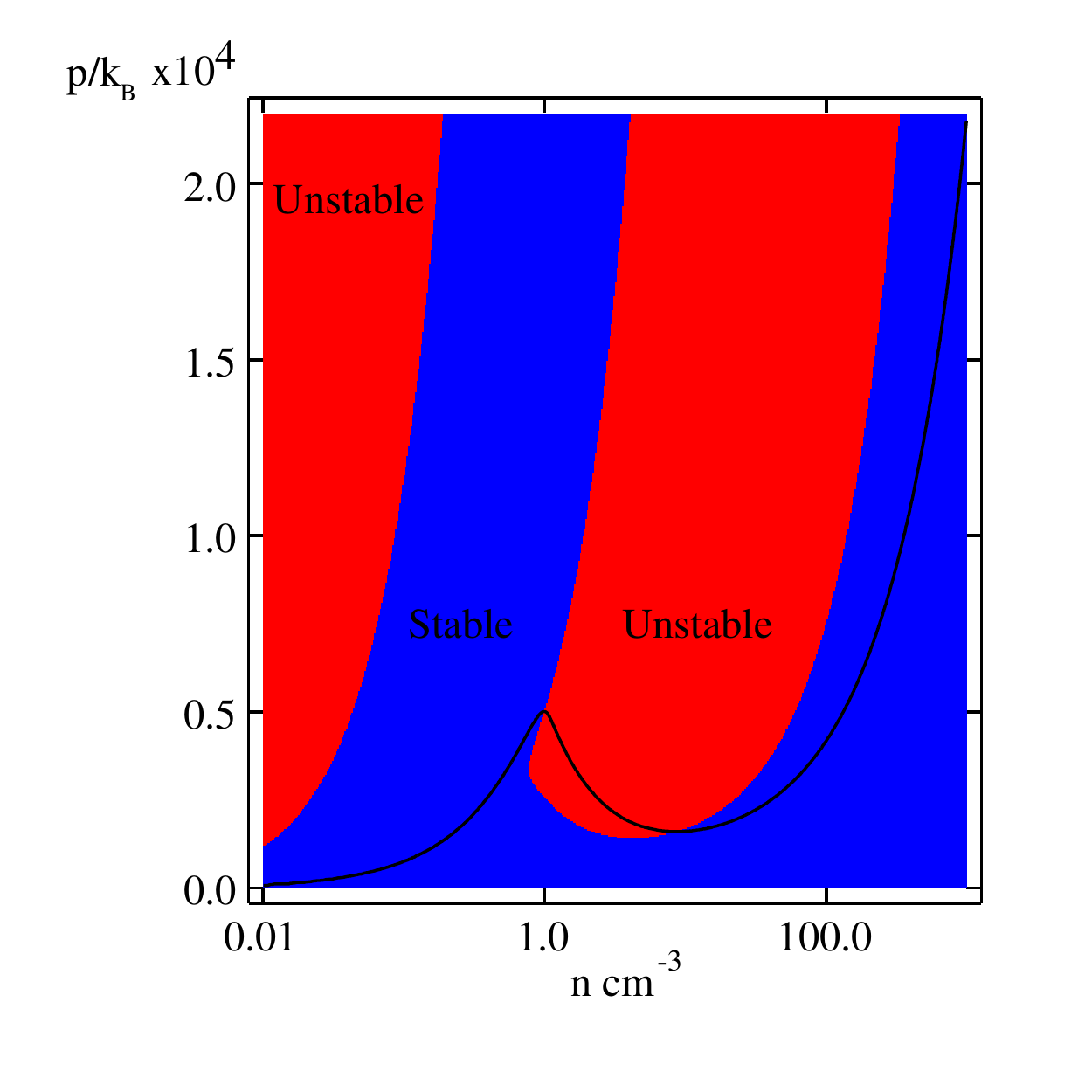}
\caption{The  unstable region  in  the $n-p$  plane  according to  the
  Balbus criterion (\ref{isoinstabcon}).  The  line is the equilibrium
  curve}
\label{fig6}
\end{figure}

\begin{figure}
\includegraphics*[viewport=0 10 375 370,width=1.0\columnwidth]{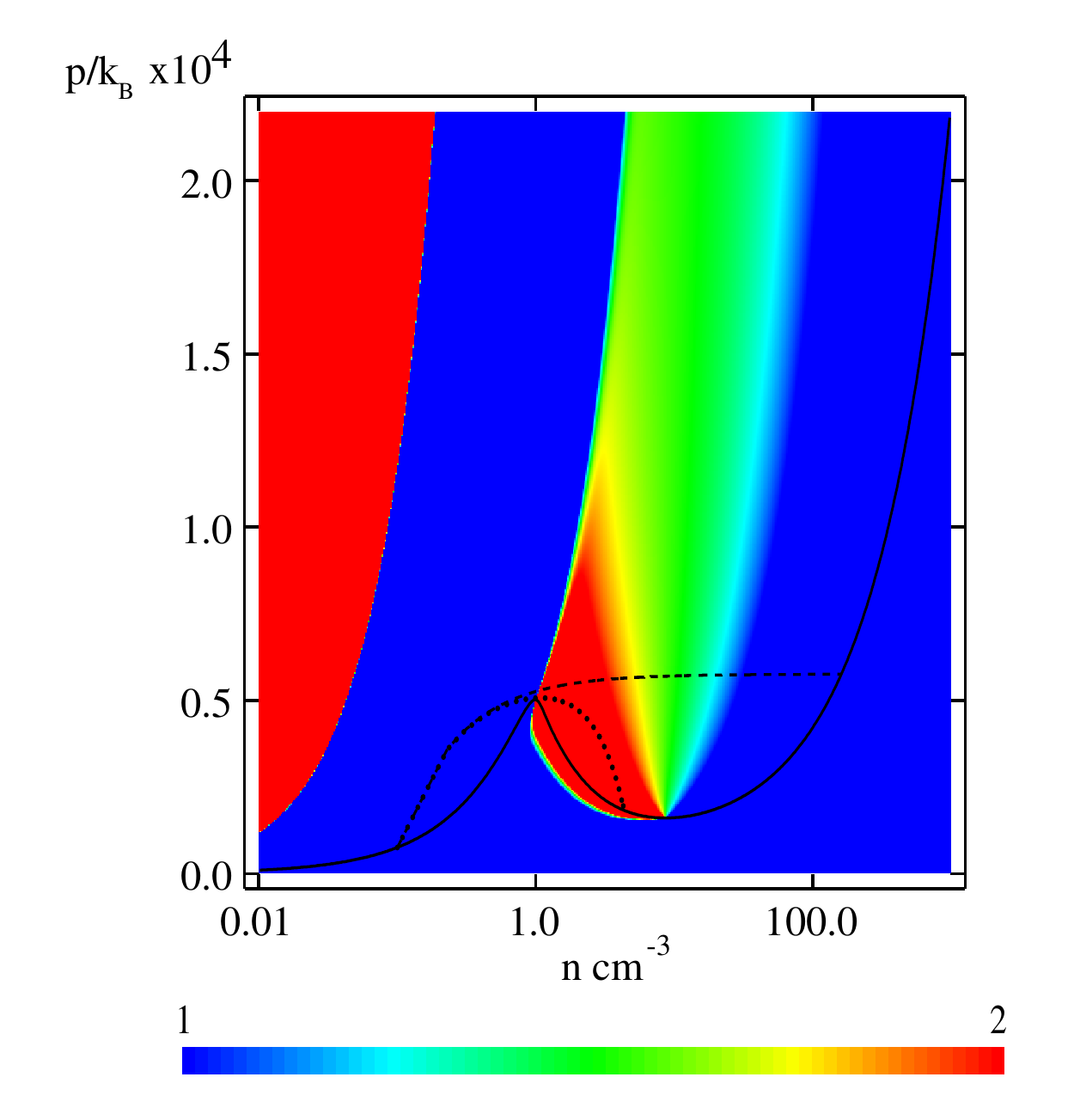}
\caption{Linear plot of the ratio of  the maximum growth rate given by
  (\ref{isolargek})   to  $|\sigma_c|$.    The  solid   line  is   the
  equilibrium  curve.  The  dashed line  is the  track of  gas passing
  through a Mach 2 hydrodynamic  shock propagating into gas in thermal
  equilibrium with  $n = 0.1$.   The dotted line  is the track  for an
  oblique MHD fast shock with a thermal Mach number of $2$ propagating
  into  an  equilibrium state  with  $n  =  0.1$, equal  parallel  and
  perpendicular  fields and  plasma $\beta  = 200$  (shock 2  in Table
  \ref{table1}).}
\label{fig7}
\end{figure}

\begin{figure*}
\begin{tabular}{ll}
  (a) & (b)\\  
\includegraphics[width=8cm]{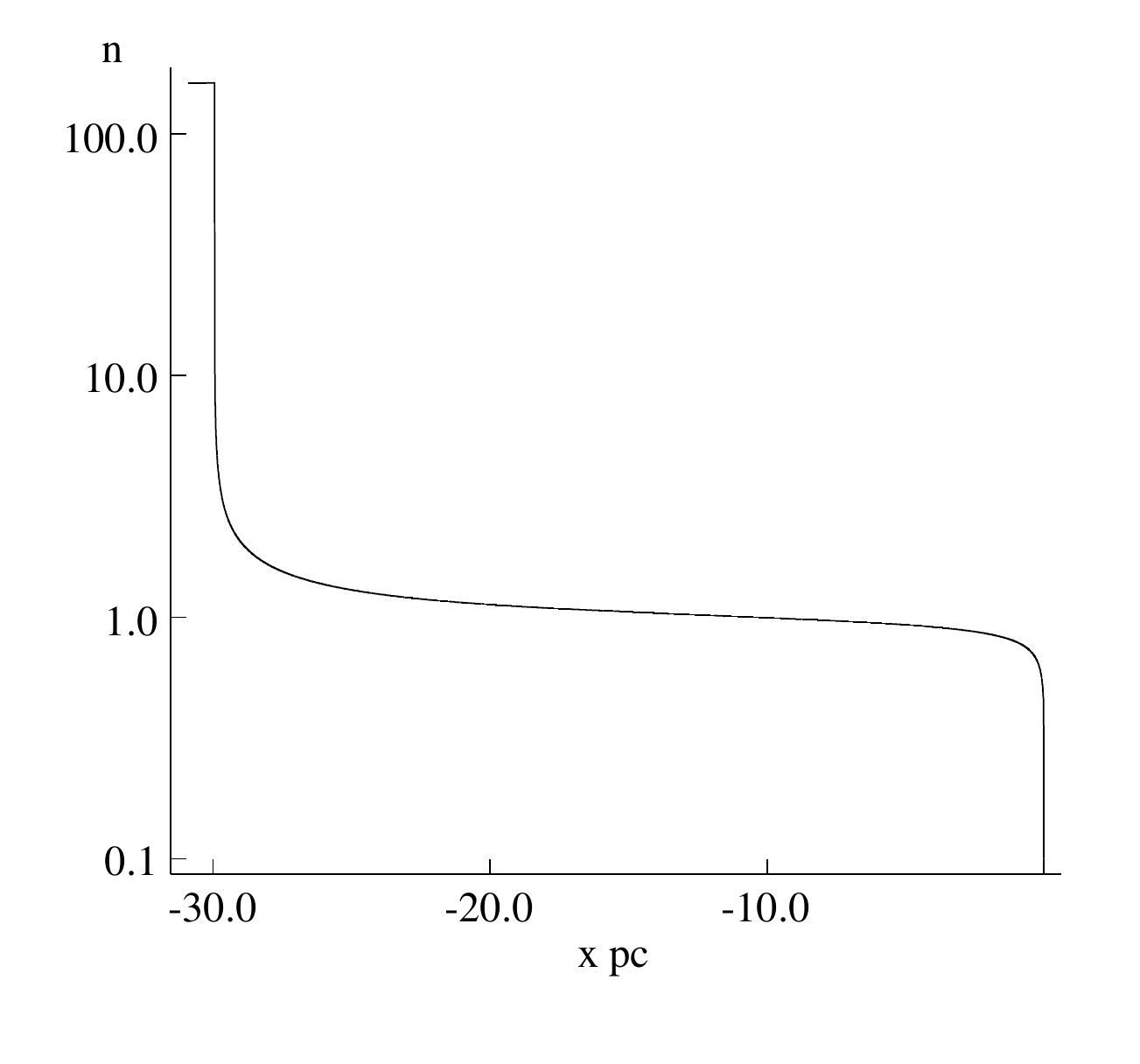}                               &
\includegraphics[width=8cm]{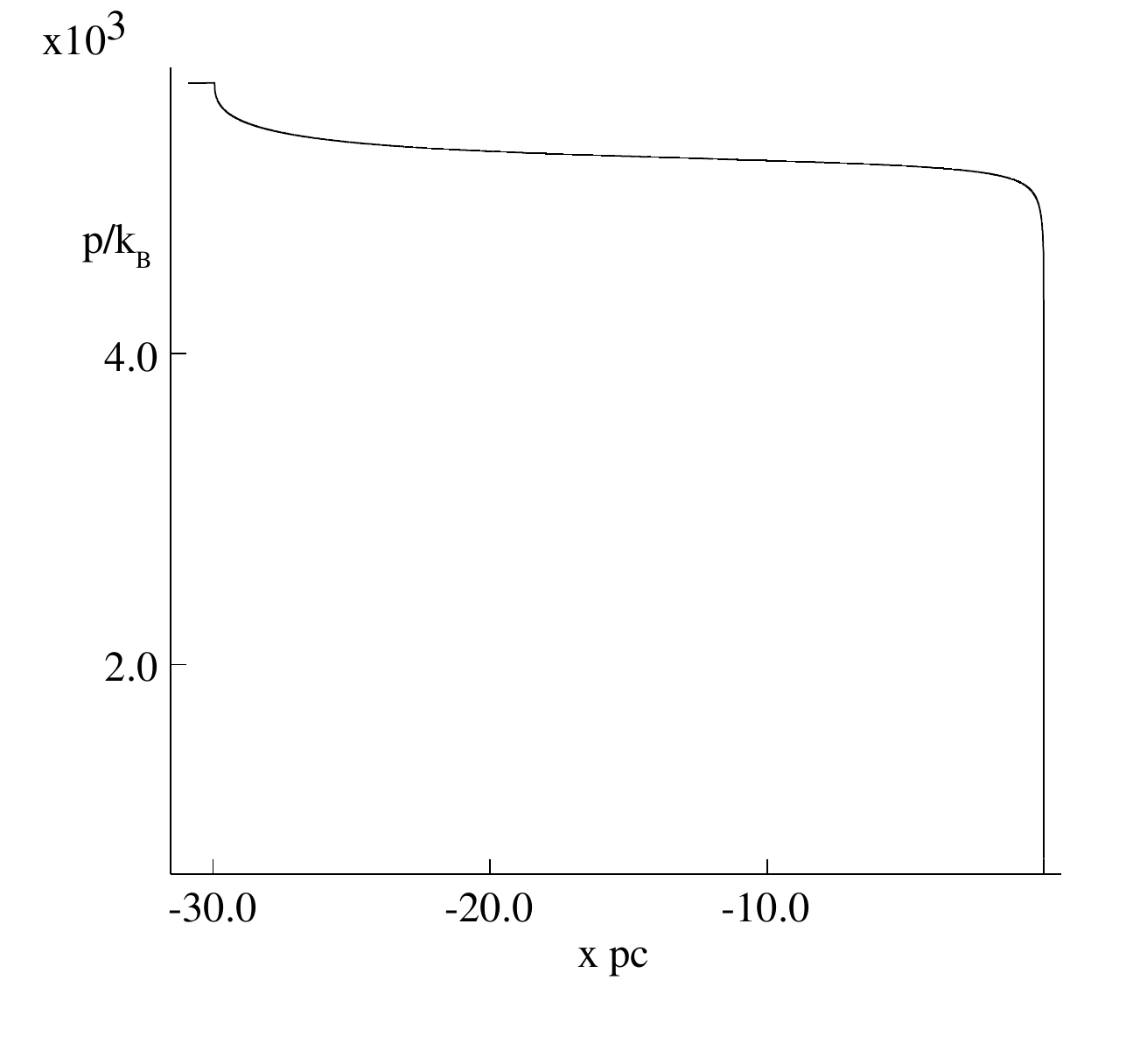} \\
  (c) & (d)\\  
\includegraphics[width=8cm]{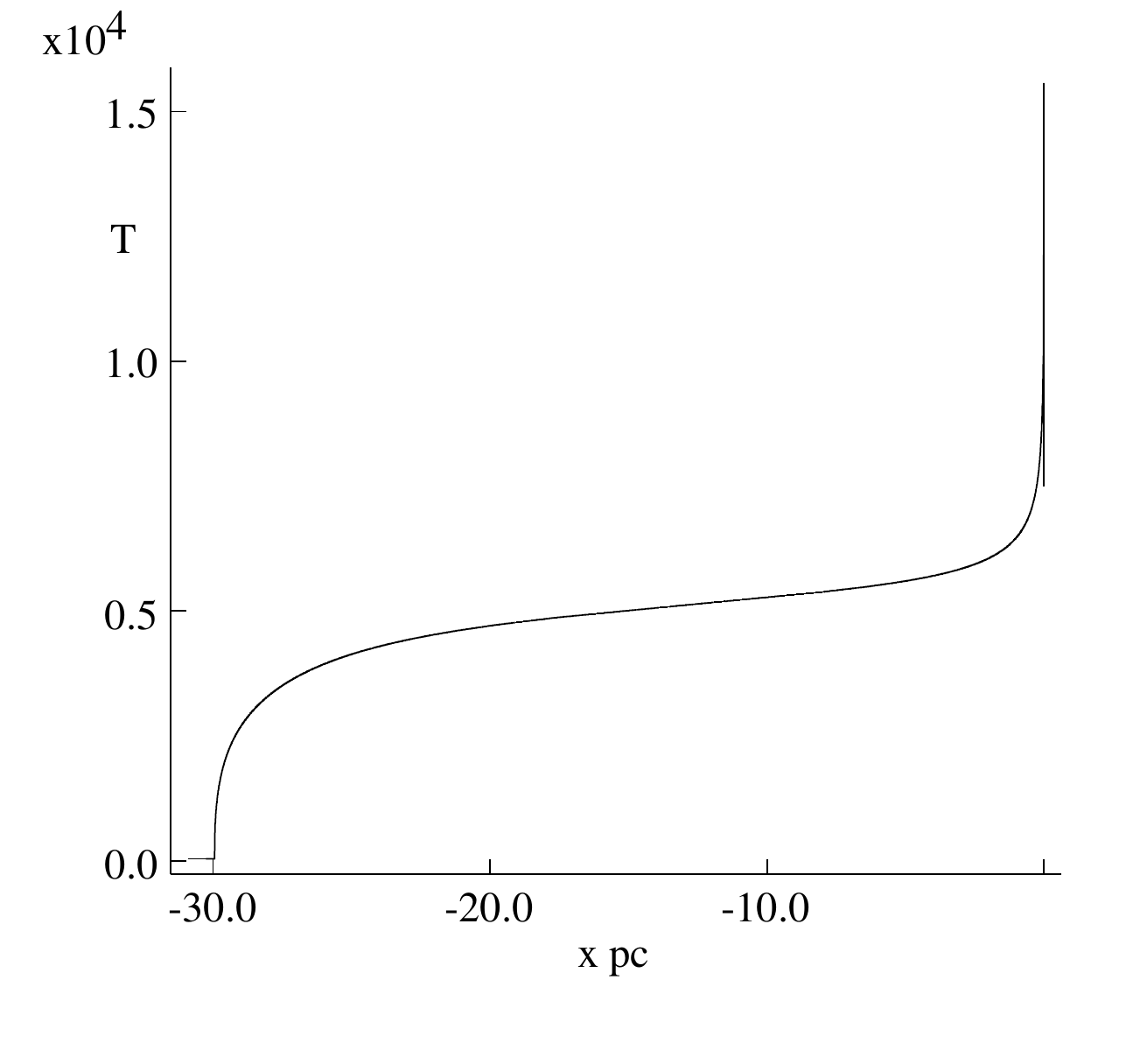}                               &
\includegraphics[width=8cm]{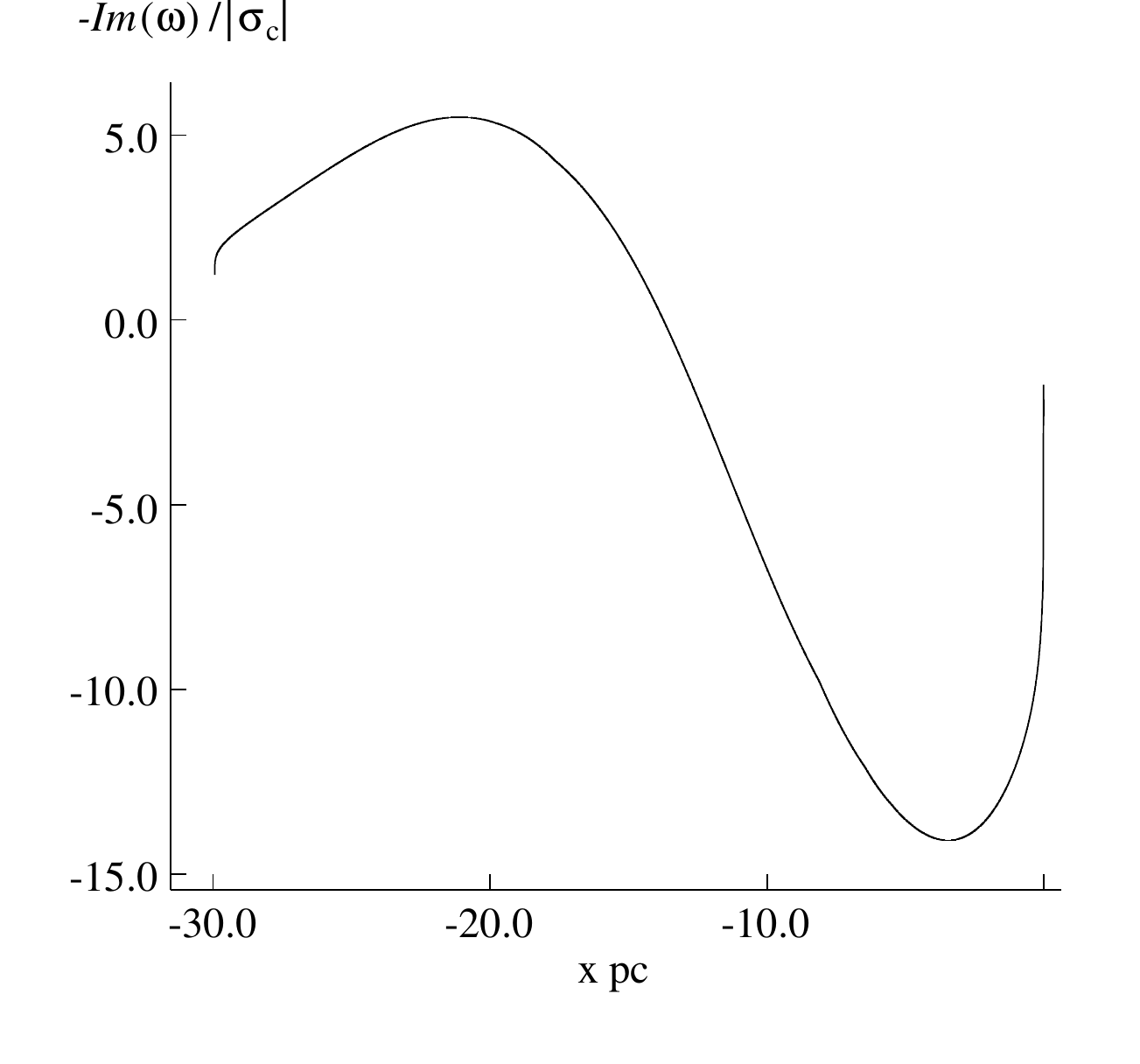}
\end{tabular}
\caption{The  solution   for  a  steady  Mach   2  hydrodynamic  shock
  propagating  into gas  in thermal  equilibrium with  $n =  0.1$.  a)
  density, b)  pressure, c) temperature,  d) the ratio of  the maximum
  growth rate  given by  (\ref{isolargek}) to $|\sigma_c|$.  Note that
  this is infinite in the final state since $|\sigma_c| = 0$ there.}
\label{fig8}
\end{figure*}

\begin{figure*}
\begin{tabular}{ll}
  (a) & (b)\\  
\includegraphics[width=8cm]{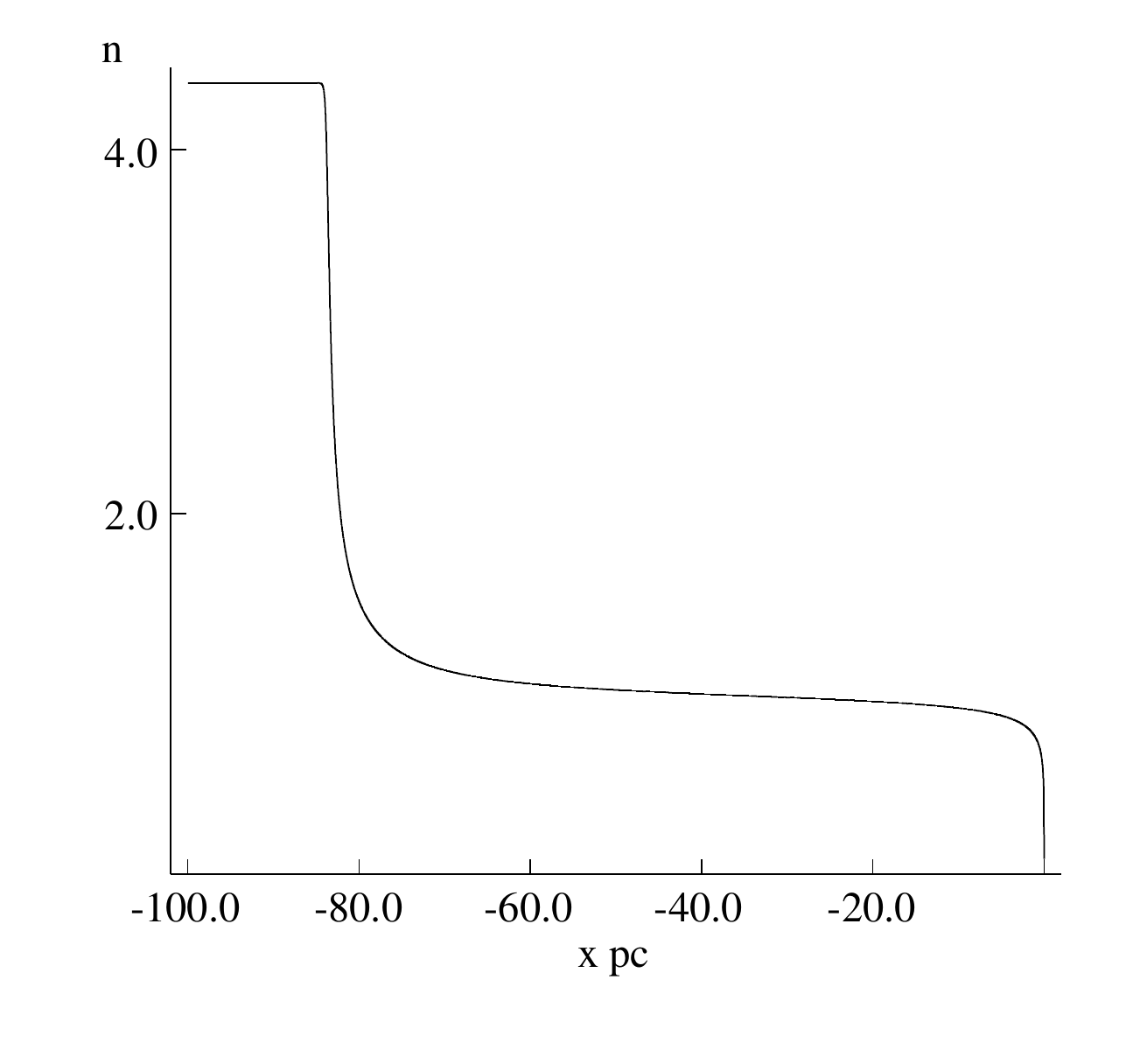}                               &
\includegraphics[width=8cm]{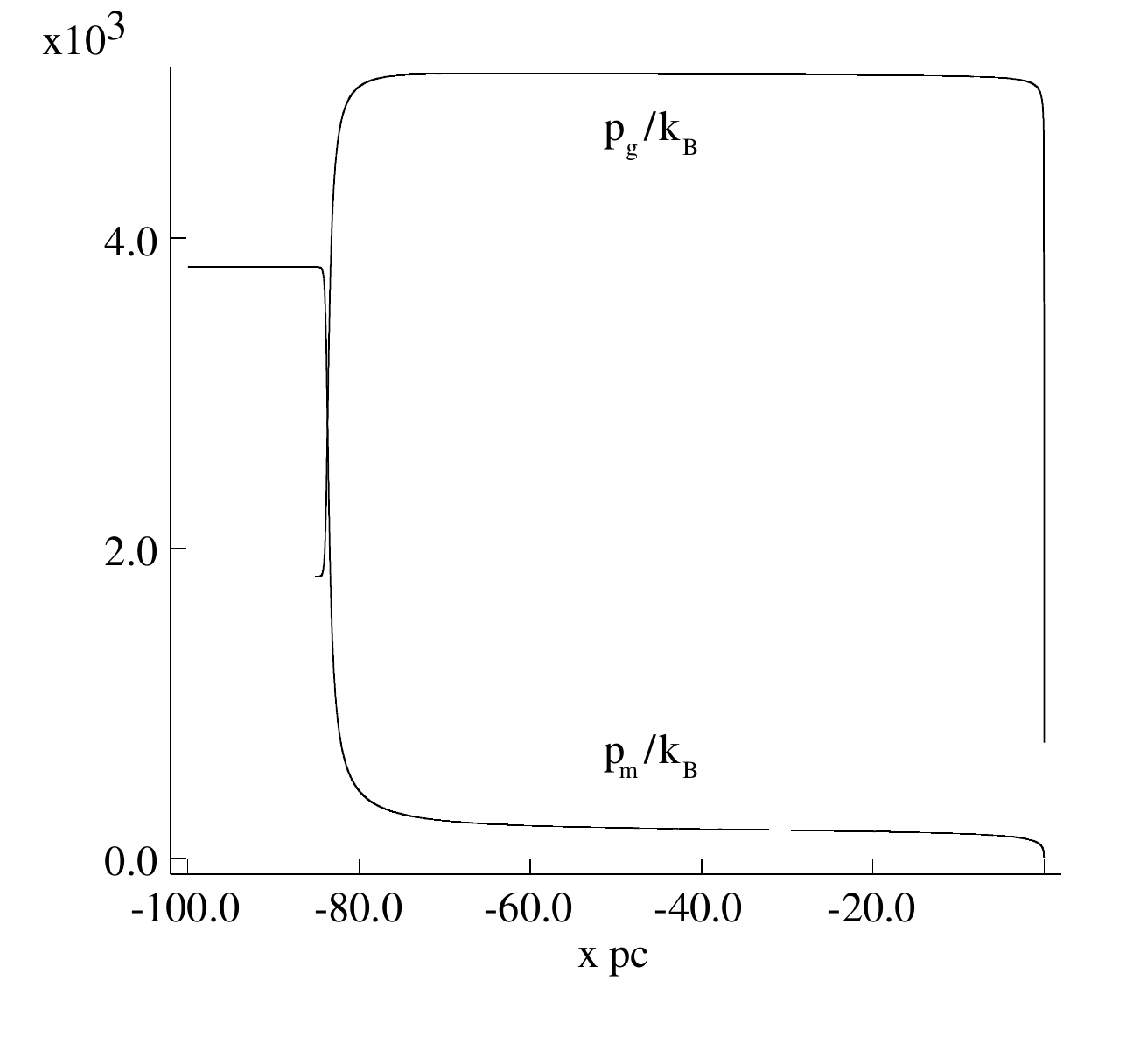} \\
  (c) & (d)\\  
\includegraphics[width=8cm]{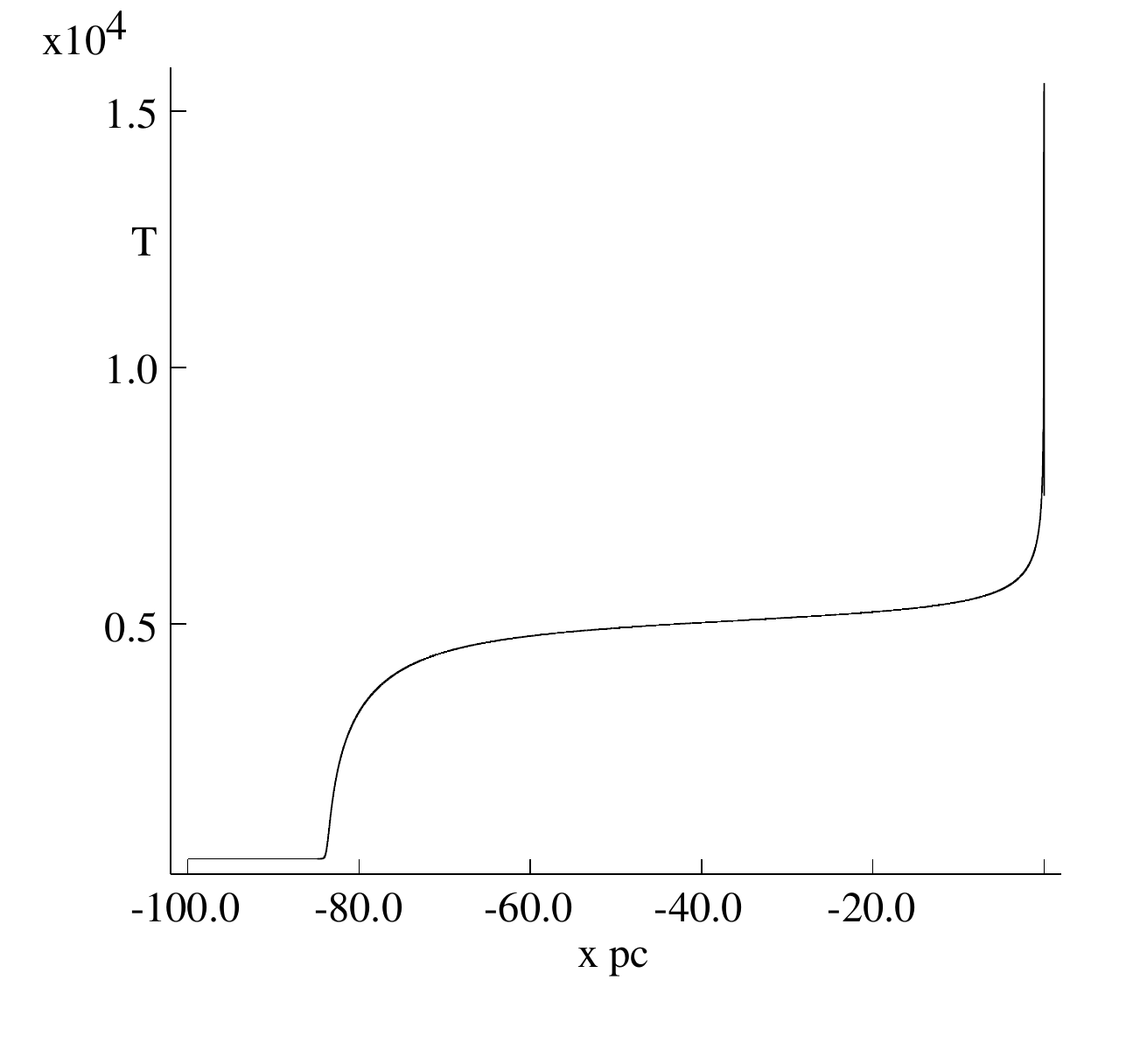}                               &
\includegraphics[width=8cm]{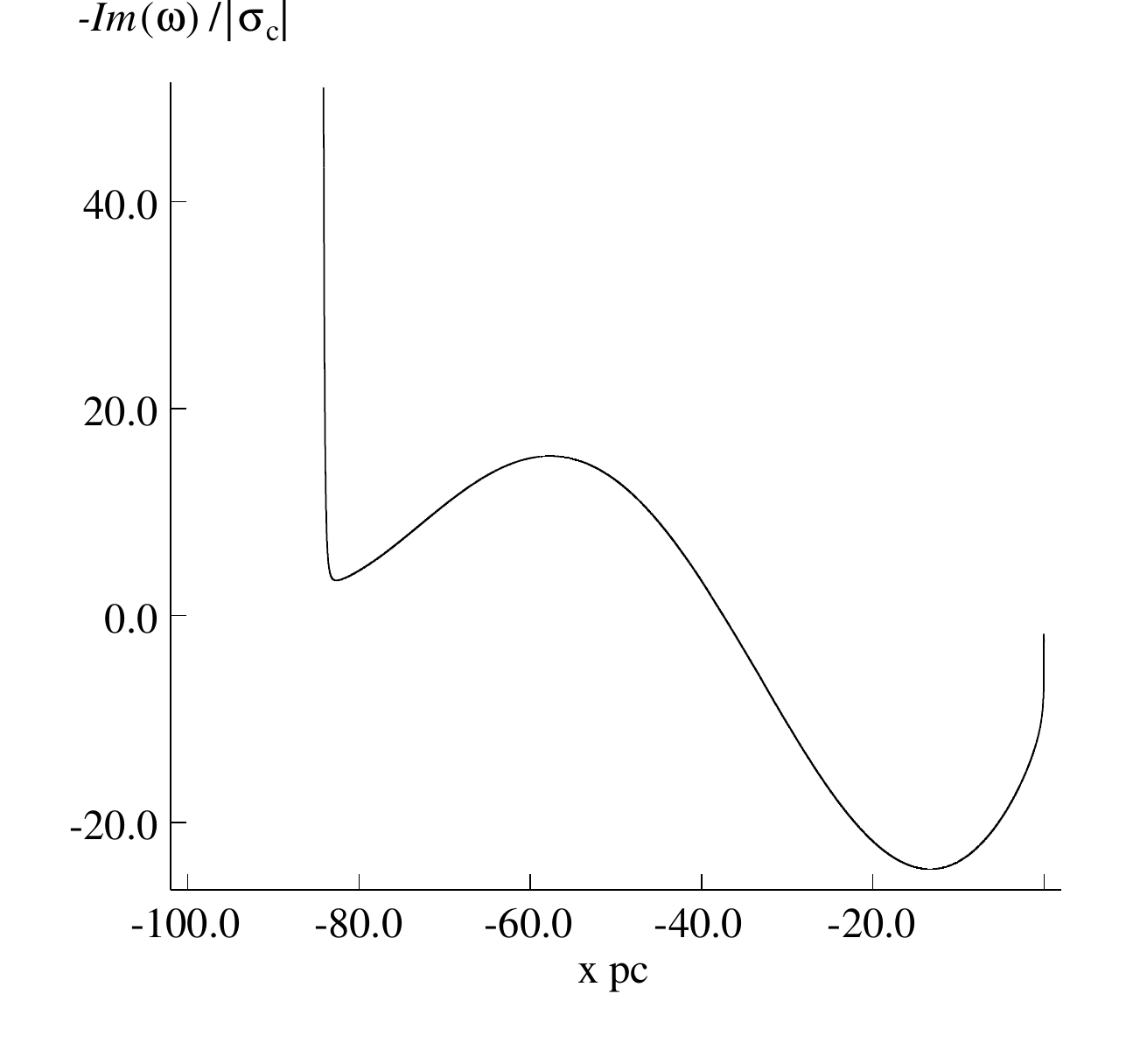}
\end{tabular}
\caption{The  solution for  a steady  oblique  MHD fast  shock with  a
  thermal Mach number of 2  propagating into an equilibrium state with
  $n = 0.1$,  $B_x = B_y$ and  plasma $\beta = 200$ (shock  2 in Table
  \ref{table1}).   a)  density,  b)  gas  and  magnetic  pressure,  c)
  temperature,  d) the  ratio  of  the maximum  growth  rate given  by
  (\ref{isolargek}) to  $|\sigma_c|$.  Note  that this is  infinite in
  the final state since $|\sigma_c| = 0$ there.}
\label{fig9}
\end{figure*}

\begin{figure*}
\begin{tabular}{l}    
(a)\\
\includegraphics*[width=17cm]{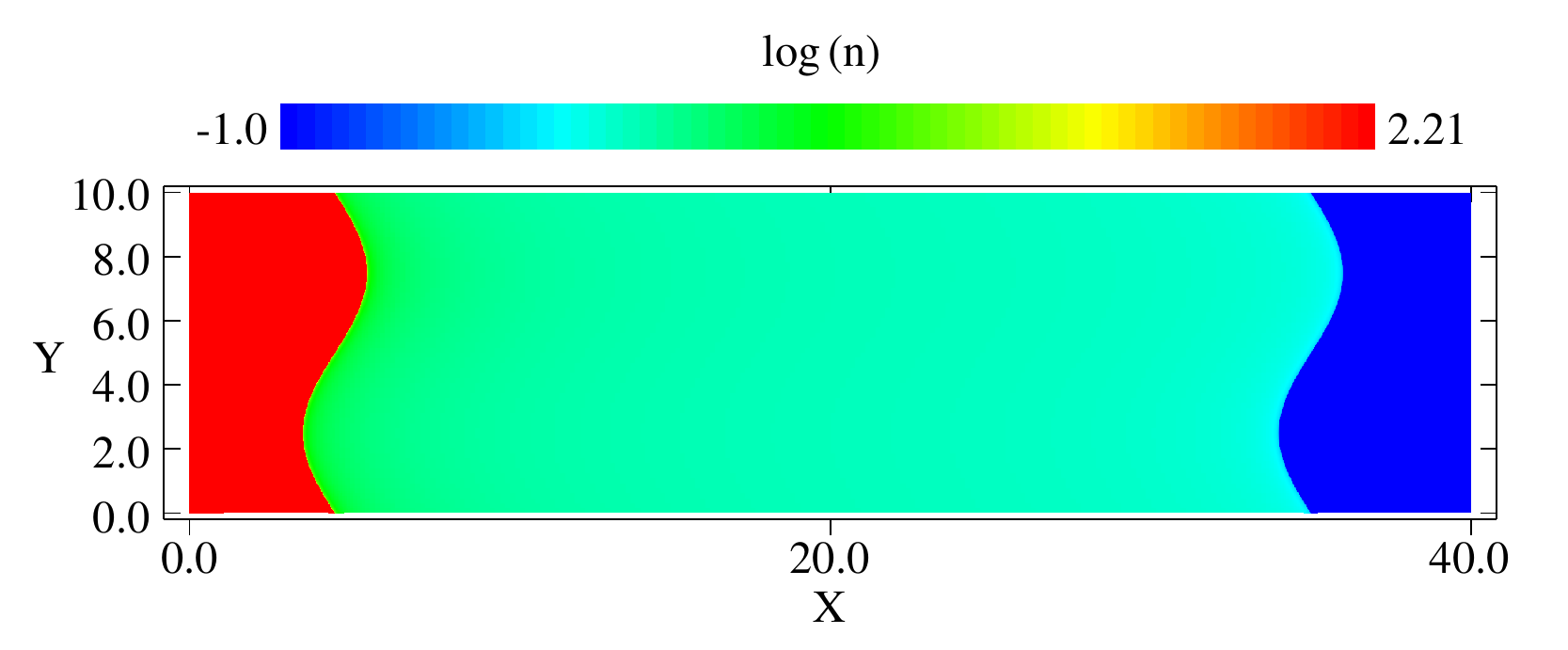}\\
(b)\\
\includegraphics*[width=17cm]{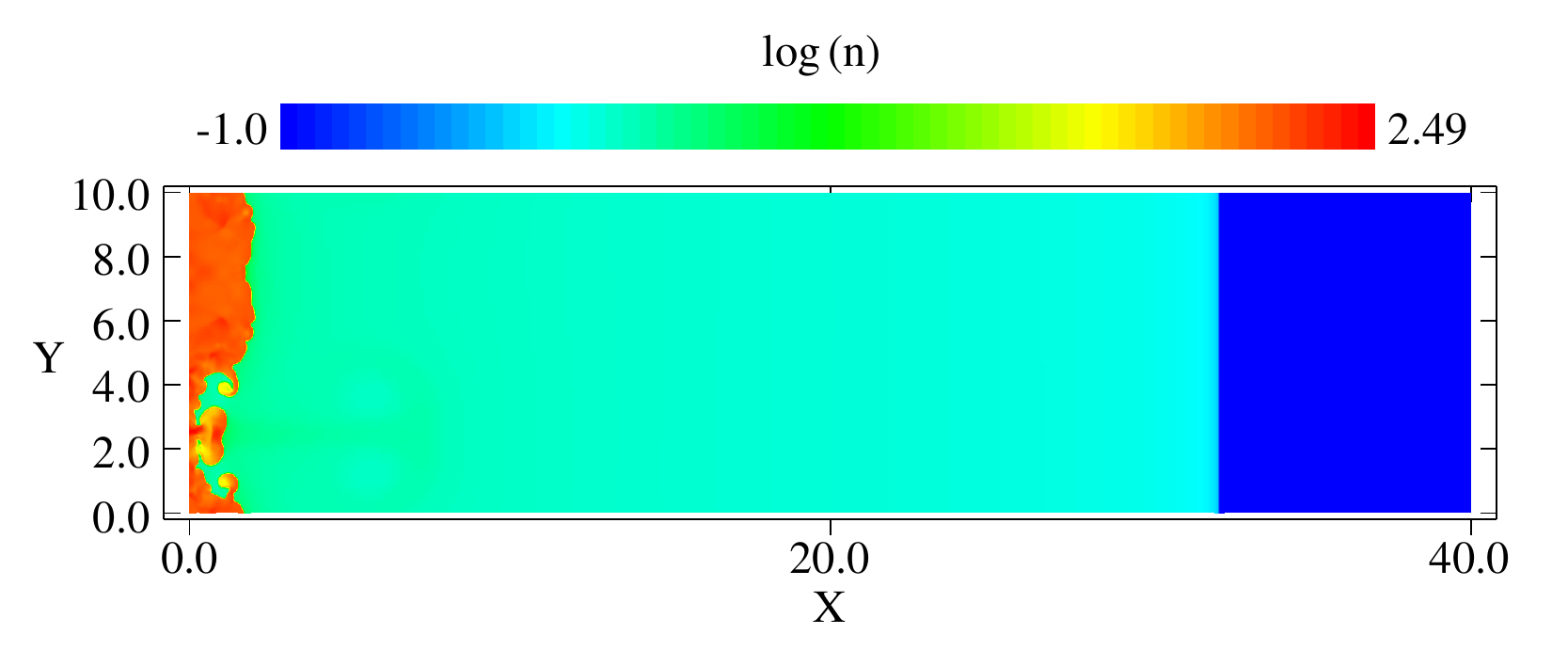}\\
(c)\\
\includegraphics*[width=17cm]{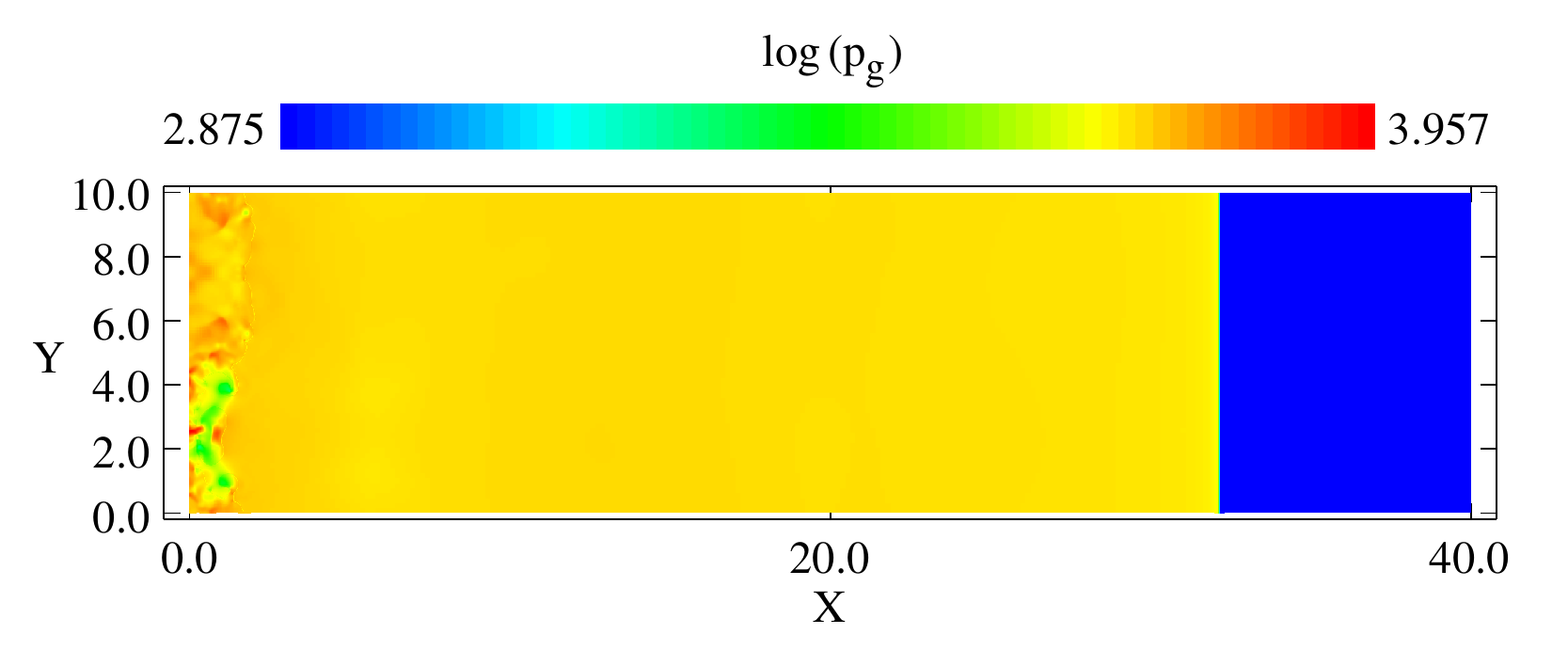}\\
\end{tabular}  
\caption{Log density  for the perturbed  2D hydrodynamic shock  1. (a)
  Initial state, (b) at  $t = 20$ Myrs.  (c) Log pressure  at $t = 20$
  Myrs.  There  were $5$  grid levels  with a  finest grid  spacing of
  $0.02$ pc. Distances are in pc.}
\label{fig10}
\end{figure*}

\begin{figure*}
\begin{tabular}{l}  
(a)\\
  \includegraphics*[width=17cm]{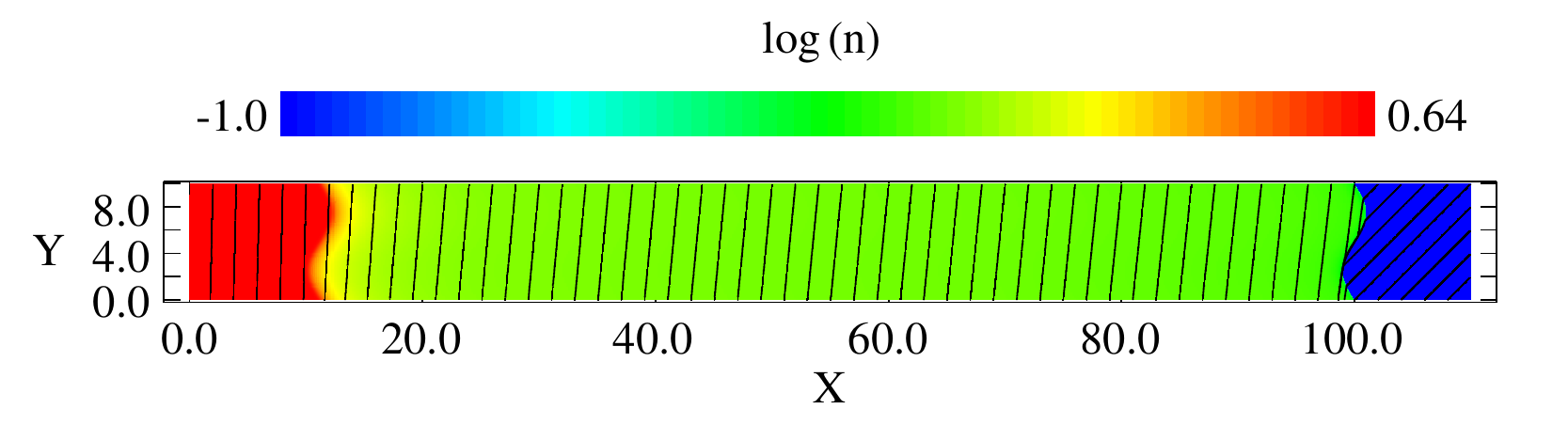}\\
(b)\\  
  \includegraphics*[width=17cm]{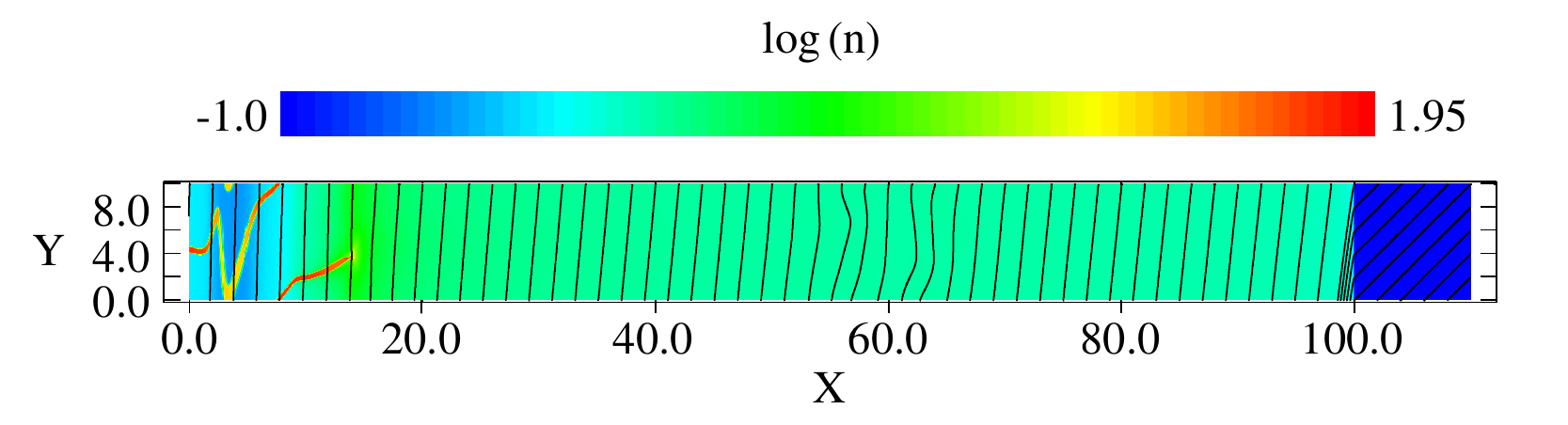}\\
(c)\\  
\includegraphics*[width=17cm]{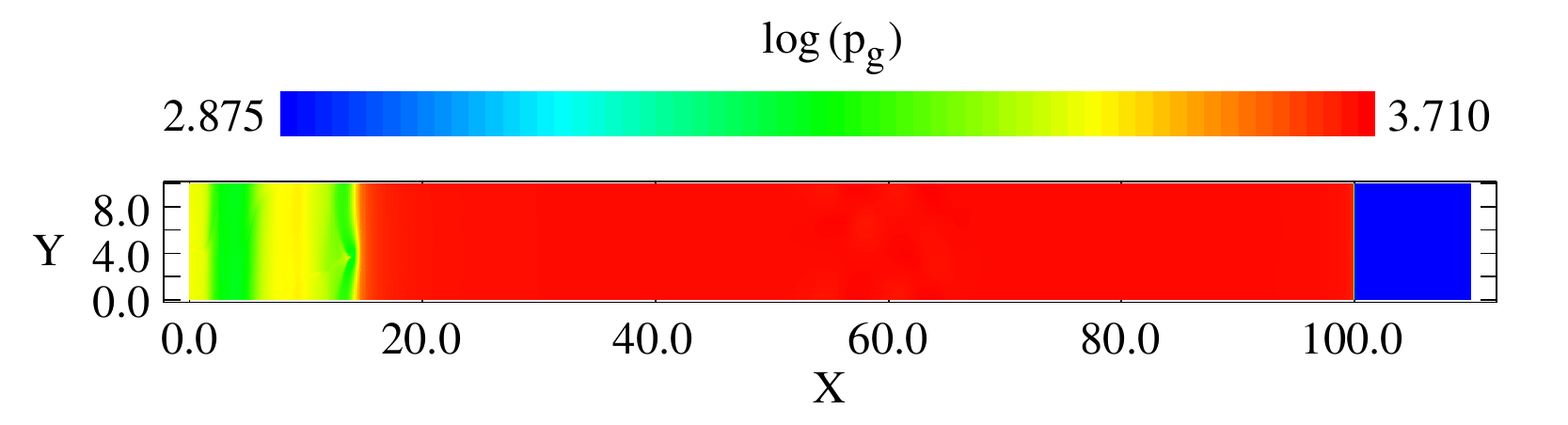}\\
\end{tabular}    
\caption{Log density and magnetic field lines for the perturbed 2D MHD
  fast shock  2 (a)  Initial state,  (b) at  $t =  20$ Myrs.   (c) Log
  pressure at $t = 20$ Myrs.  There were $5$ grid levels with a finest
  grid spacing of  $0.02$ pc. Distances are in pc.   The FWHM width of
  the filaments is $\simeq 0.52$ pc.}
\label{fig11}
\end{figure*}

\begin{figure*}
  \begin{tabular}{l}
(a)\\
  \includegraphics*[width=17cm]{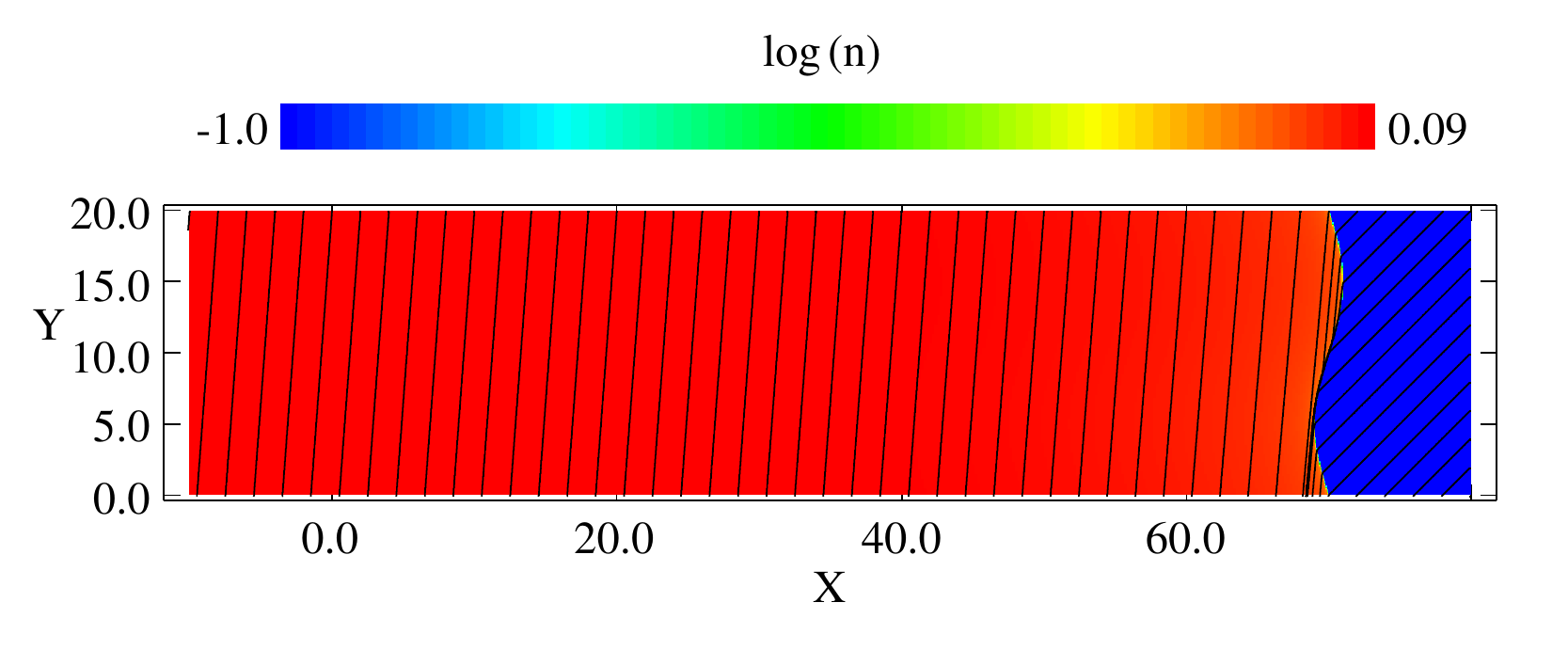}\\
(b)\\  
  \includegraphics*[width=17cm]{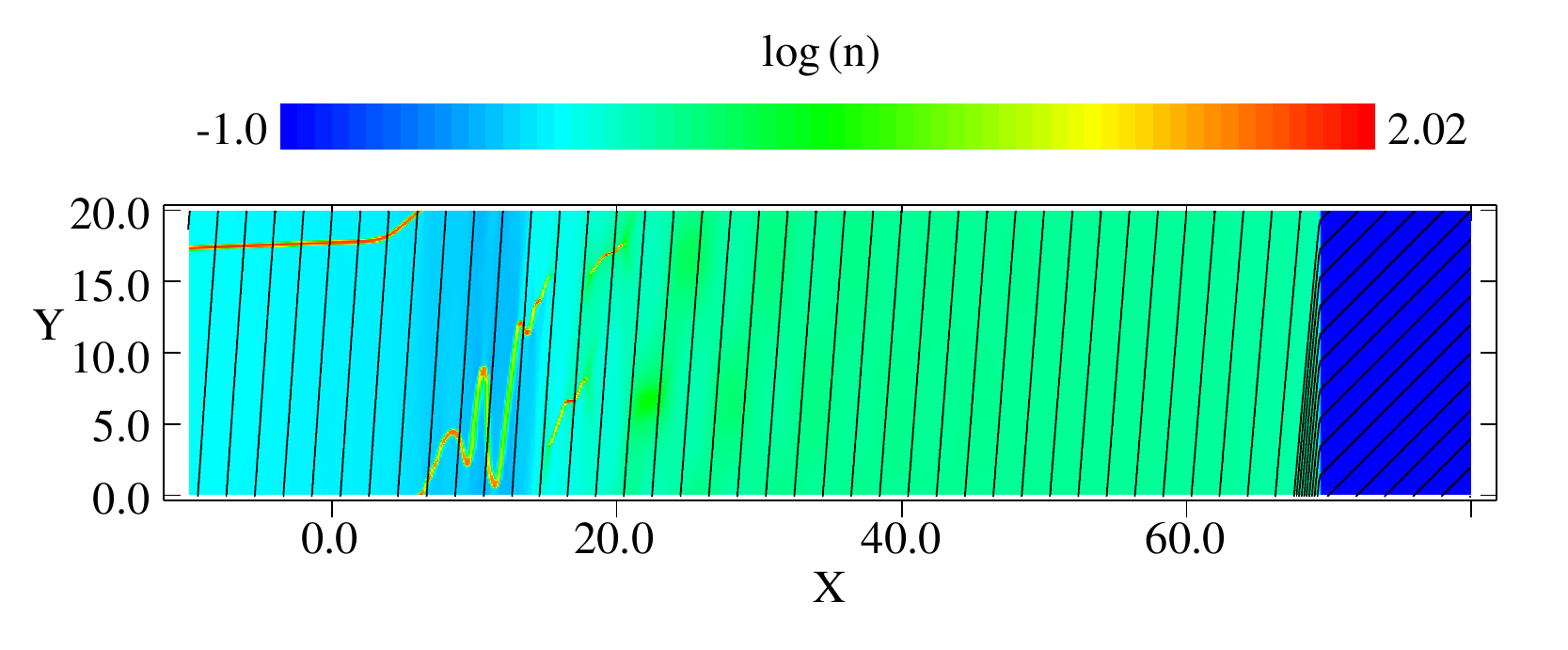}\\
(c)\\  
\includegraphics*[width=17cm]{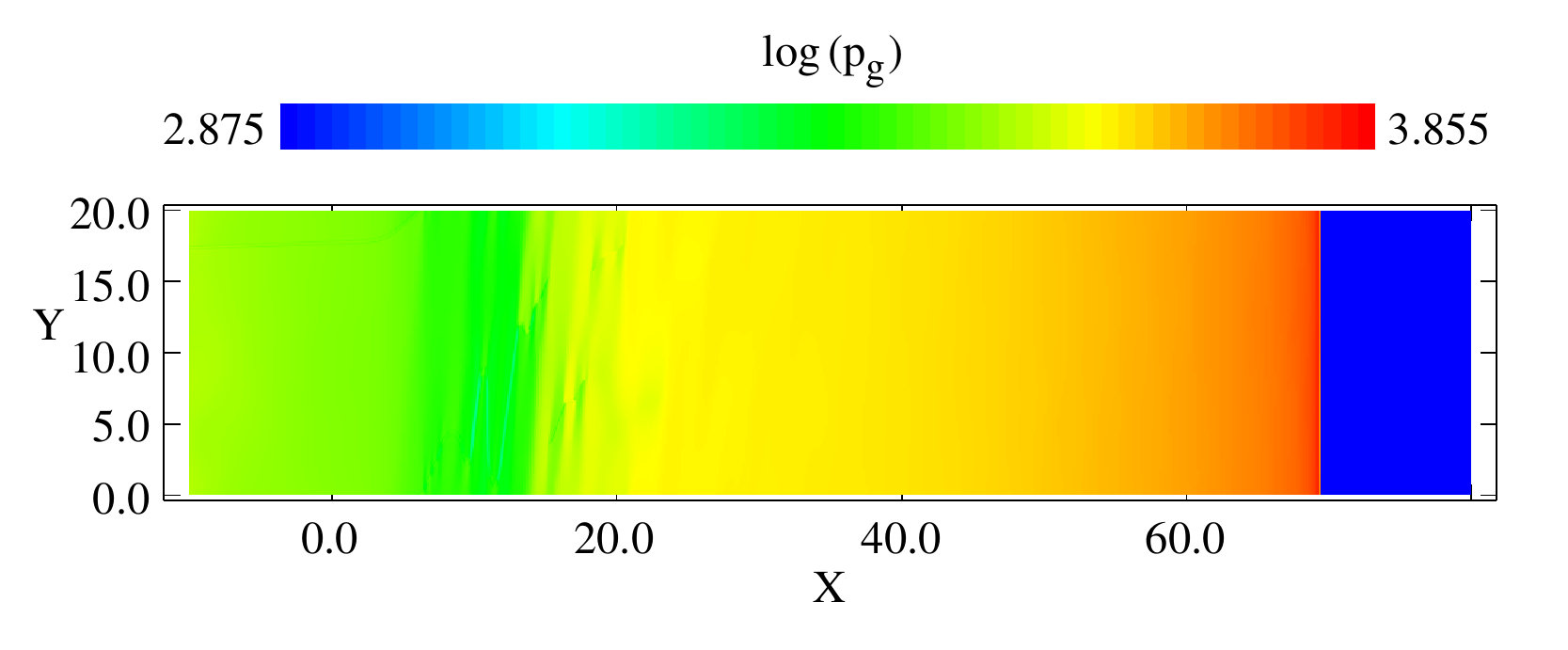}\\    
\end{tabular}    
\caption{Log density and magnetic field lines for the perturbed 2D MHD
  fast shock  5 (a)  Initial state,  (b) at  $t =  25$ Myrs.   (c) Log
  pressure at $t = 25$ Myrs.  There were $5$ grid levels with a finest
  grid spacing of  $0.02$ pc. Distances are in pc.   The FWHM width of
  the filaments is $\simeq 0.52$ pc.}
\label{fig12}
\end{figure*}

\begin{figure}
\includegraphics*[width=8cm]{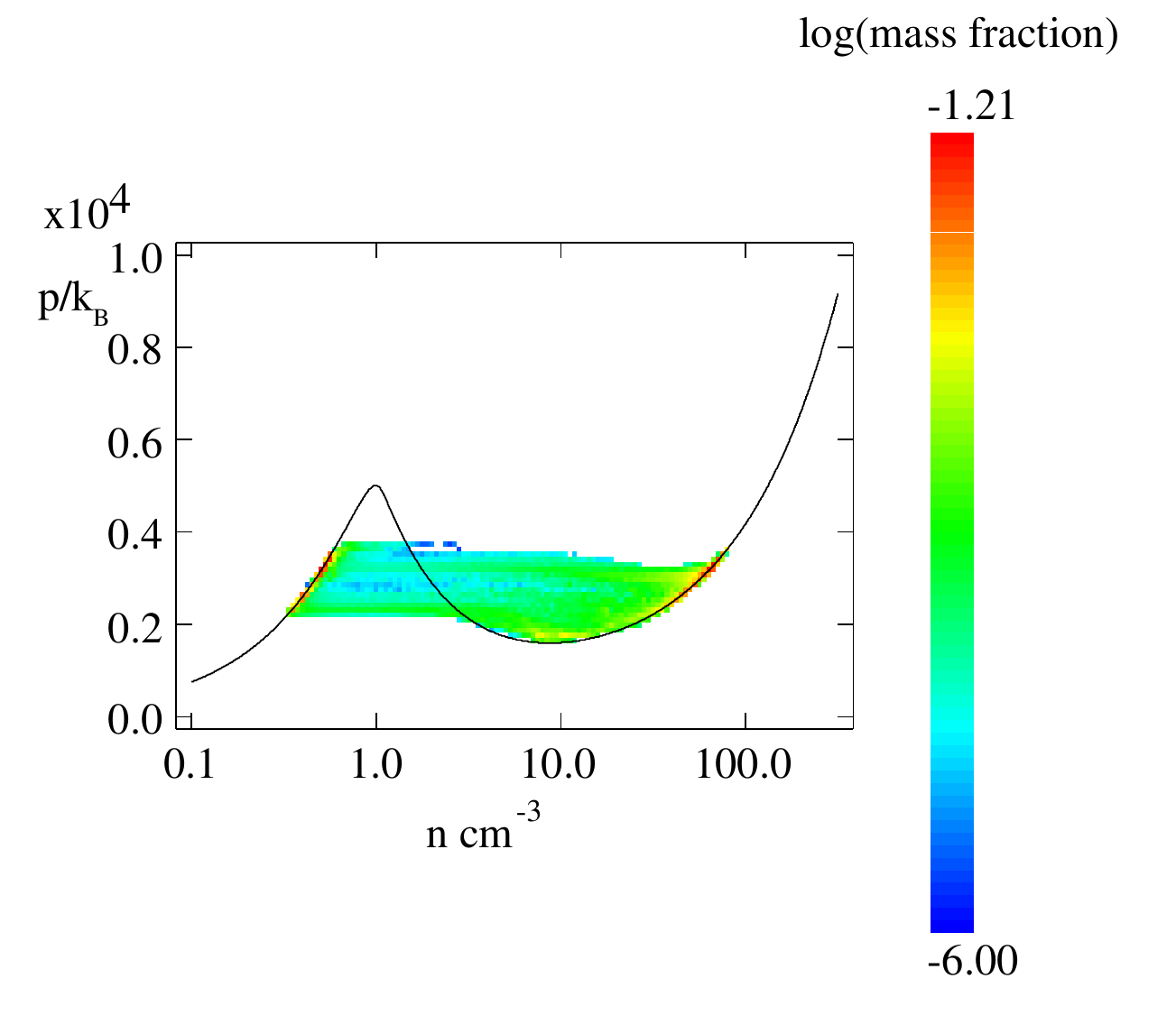}
\caption{Distribution  of mass  fraction  in the  $n-p$  plane in  the
  region $-10 \le x  \le 15$ for the perturbed 2D MHD  fast shock 5 at
  $t = 25$ Myrs.  The integrated mass fractions are: $0.3589$ for warm
  gas, $0.5748$ for cold gas and $0.066$ for unstable gas.}
\label{fig13}
\end{figure}

\begin{figure}
\begin{tabular}{l}
(a)\\
  \includegraphics[width=8cm]{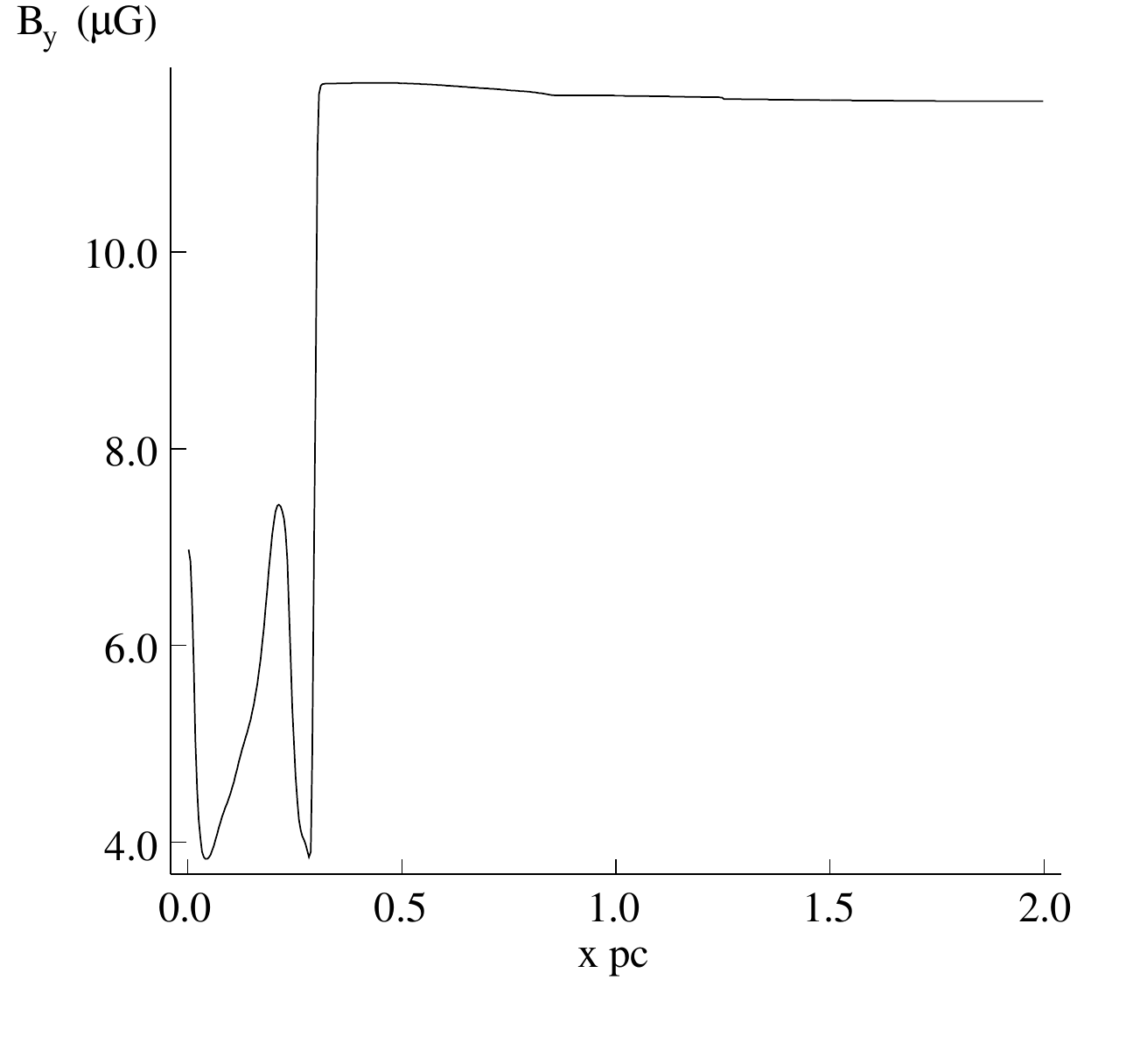} \\
(b)\\
  \includegraphics[width=8cm]{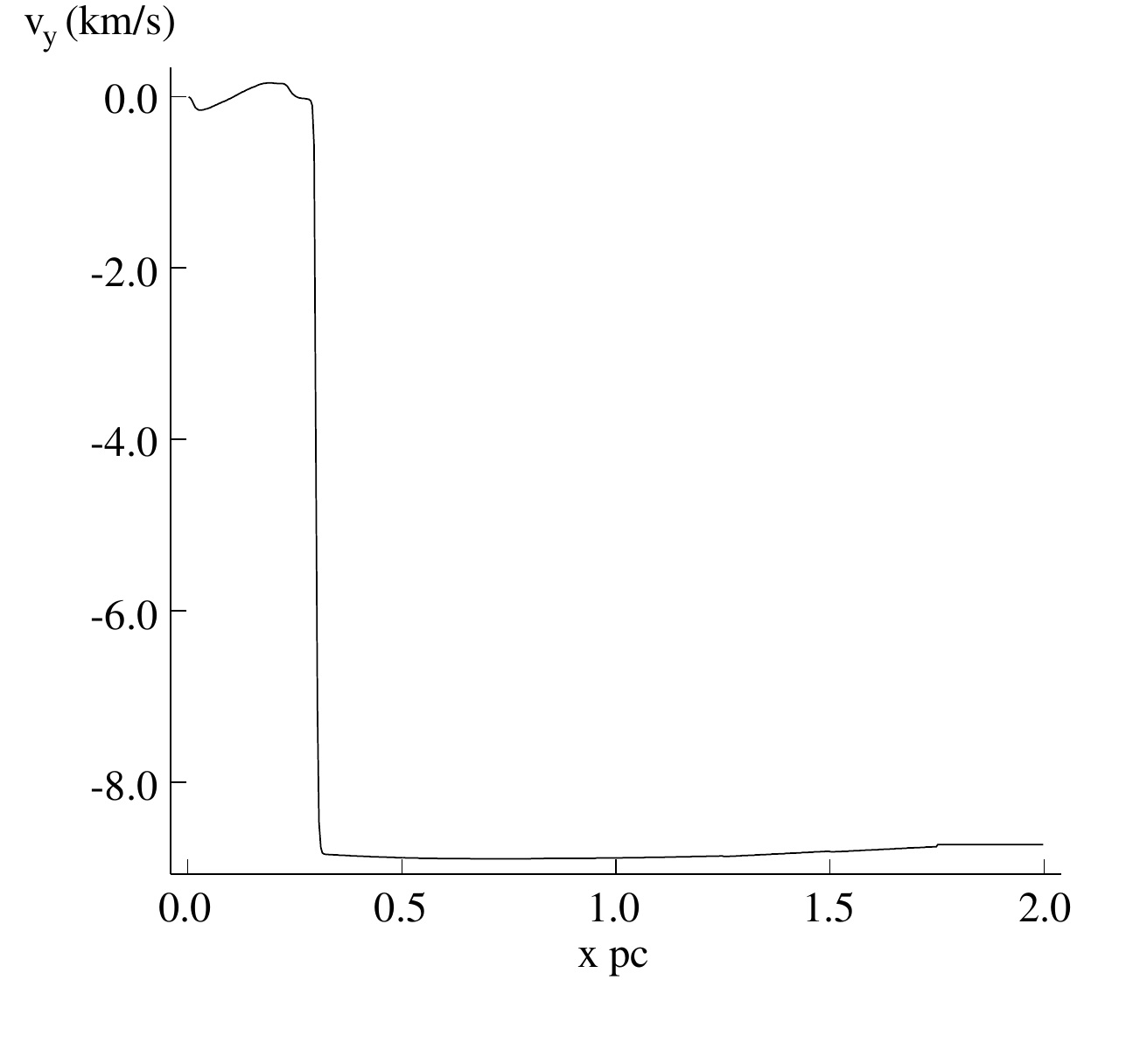}    
\end{tabular}
\caption{Region near the  interface for \protect\cite{Inoue:2009} case
  1a (shock 11 in Table \ref{table1})  at $t = 60$ Myr.  a) transverse
  magnetic field, b) transverse velocity.  At this time the fast shock
  is at $x = 88.075$.}
\label{fig14}
\end{figure}

\begin{figure*}
\includegraphics*[width=18cm]{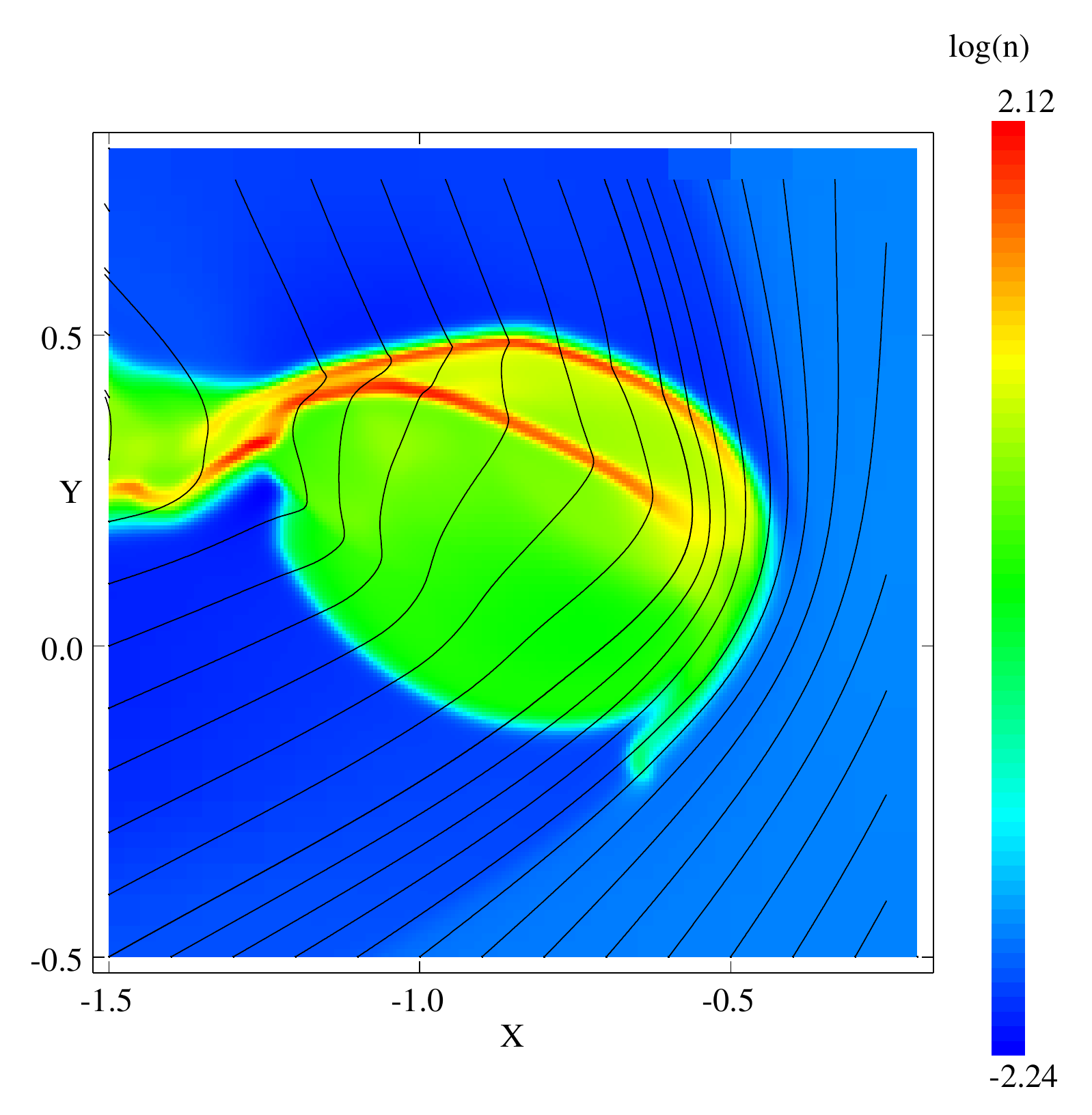}
\caption{Log  density and  magnetic  field lines  for the  shock-cloud
  interaction at $t = 14.5$ Myrs.  Distance is in units of the initial
  cloud radius ($200$ pc).}
\label{fig15}
\end{figure*}

\section{Koyama \& Inutsuka energy source}
\label{Koyama_en}

\cite{Koyama:2002} used a thermal energy loss function of the form

\vspace*{-10pt}
\begin{equation}
L(\rho, T) = \frac{\rho}{m_H^2} \Lambda (T) - \frac{1}{m_H} \Gamma,
\label{energysource}
\end{equation}
\vspace*{-5pt}

\noindent
where $T$ is in Kelvin, $\Gamma = 2~10^{-26}$ erg s$^{-1}$ and

\vspace*{-10pt}
\begin{equation}
\frac{\Lambda}{\Gamma}  =   10^7  \exp  \left(   \frac{-1.184~  \times
  ~10^5}{T  +   1000}  \right)  +  1.4~10^{-2}   T^{1/2}  \exp  \left(
\frac{-92}{T} \right),
\label{cool}
\end{equation}
\vspace*{-5pt}

\noindent
This    has   been    used   by    a   number    of   authors    (e.g.
\citealt{Vazquez-Semadini:2007,        Kim:2008,        Wareing:2016a,
  Wareing:2016b, Wareing:2017, Wareing:2018, Wareing:2019}).

\cite{Koyama:2004} include a thermal conductivity

\vspace*{-10pt}
\begin{equation}
\kappa = 2.5~10^3~T^{1/2},
\label{conduct}
\end{equation}
\vspace*{-5pt}

\noindent
which is appropriate for $T < 4.5~10^4~K$ (\citealt{Parker:1953}). The
kinematic viscosity is given by

\vspace*{-10pt}
\begin{equation}
\nu = \frac{(\gamma - 1)}{\gamma} \frac{m}{k_B} \frac{\kappa}{\rho} P_r,
\label{kinvis}
\end{equation}
\vspace*{-5pt}

\noindent
where $P_r$ is the Prandtl number, which is $2/3$ for a monotomic gas.

\subsection{Equilibrium states}

The  unstable   temperature  and  density  ranges   for  the  isobaric
condensation mode are $184$ K $\le T \le 5039$~K and $0.9936 \le n \le
8.6818$.  The equilibrium pressure is shown in Fig.  \ref{fig1}.  From
Fig.  \ref{fig2}  we can  see that  $\alpha <  1$ everywhere,  so that
equation (\ref{stabcon}) tells us that the isentropic modes are always
stable.   Fig.   \ref{fig3}  shows  the Field  length  from  equations
(\ref{fieldlambda}) and  (\ref{koyamalambda}) in the  unstable region,
from which it can be seen that  they are indeed very different. We can
see from Fig.  \ref{fig4} that  the thermal wavelength is rather large
in most  of the unstable region,  varying between the typical  size of
giant molecular clouds  and that of the  translucent clumps.  Although
Fig. \ref{fig5} shows that the growth rate of the condensation mode as
a function of wavelength does have a  maximum, it is not so sharp that
one particular wavelength is strongly favoured.

\subsection{Non-equilibrium states}

A  number  of authors  have  considered  instability occurring  behind
shocks  generated  by colliding  flows  that  drive  the gas  into  an
unstable   non-equilibrium   state  (e.g.    \citealt{Hennebelle:2000,
  Koyama:2000,  Koyama:2002,   Heitsch:2009,  Inoue:2008,  Inoue:2009,
  Fogerty:2016}).   Fig.  \ref{fig6}  shows  the region  in the  $n-p$
plane in  which (\ref{isoinstabcon}) indicates instability  and agrees
with fig.   4 in \cite{Inoue:2008}.  Note that the unstable  region at
low densities is unphysical since it corresponds to temperatures above
$10^4$ K, for which the energy source function (\ref{energysource}) is
not valid.  A  more realistic model of the  interstellar cooling curve
above  $10^4$ K,  such as  that in  \cite{Gnat:2012}, gives  isochoric
instability for $T > 10^5$ K.

\cite{Koyama:2000} applied  the analysis in  Section \ref{Nonequistab}
to thermally unstable gas cooling behind a shock, but it is only valid
when the growth  rate given by (\ref{isolargek}) is  large compared to
the  rate  of contraction,  $|\sigma_c|  =  |\dot{R}/R|$. Gas  cooling
behind shocks that lead to a phase change must indeed pass through the
unstable  region above  the  equilibrium curve,  but Fig.   \ref{fig7}
shows  that  the  maximum  growth   rate  is  not  large  compared  to
$|\sigma_c|$ in most of this region.   In fact the analysis only gives
reasonable quantitative results for shocks  that are not much stronger
than that required to trigger a transition to the cold phase.

\begin{table*}
\begin{center}
\caption{Properties of steady shock solutions: $M_{Th}$ - thermal Mach
  number; $\theta$ - angle between  field and shock perpendicular; $n$
  - particle number density; $\beta$ - plasma $\beta$; $B$ - magnitude
  of  the magnetic  field; $l_c$  - length  of cooling  region; $A$  -
  amplification factor given by  (\ref{ampfac}).  The suffices $u$ and
  $f$  denote  the  upstream  and  final  values.   Shock  10  is  the
  perpendicular    shock    generated     by    the    collision    in
  \protect\cite{Inoue:2008} and shock 11  the case 1a $15^\circ$ shock
  in \protect\cite{Inoue:2008} }
\label{table1}
\begin{tabular}{l|l|l|l|l|l|l|l|l|l|l|l|}
&  ~~~~~$M_{Th}$  &  $\theta_u$  &  $\theta_f$ &  $n_u$  &  ~~$n_f$  &
  $\beta_u$  &  $\beta_f$ &  $B_u  ~\mu$G  &  $B_f~\mu$G &$l_c$  pc  &
  ~~A\\ \hline\hline 
1  & $2.0~  (18.56$ km  s$^{-1}$) &  NA &   NA &  $0.1$ &  $162.5$  &
$\infty$ & $\infty$ & $0$ & $0$ & $30.49$ & $3.5~10^3$\\
2 &  $2.0~ (18.56$ km s$^{-1}$)  & $45^\circ$ & $88.7^\circ$&  $0.1$ &
$4.368$ & $200$ & $0.4764$ & $0.114$ & $3.638$ &$85.04$ & $4.0~10^4$\\
3 & $3.0~ (27.84$ km s$^{-1}$) & NA  & NA & $0.1$ & $454.1$ & $\infty$
& $\infty$ & $0$ & $0$ &$1.928$ & $2.7~10^2$\\
4 &  $3.0~ (27.84$ km s$^{-1}$)  & $45^\circ$ & $89.2^\circ$&  $0.1$ &
$7.212$  &  $200$  &  $0.1578$  &   $0.114$  &  $5.954$  &  $2.782$  &
$9.3~10^1$\\
5 & $3.0~  (27.84$ km s$^{-1}$) & $45^\circ$ &  $85.8^\circ$ & $0.1$ &
$1.243$ & $10$ & $0.6350$ & $0.510$ & $4.865$ & $66.05$ & $4.4~10^2$\\
6 & $2.0~ (17.02$ km s$^{-1}$) & NA & NA & $0.5$ & $1126$ & $\infty$ &
$\infty$ & $0$ & $0$ &$1.336$ & $2.7~10^2$\\
7 & $2.0~  (17.02$ km s$^{-1}$) & $45^\circ$ &  $85.5^\circ$ & $0.5$ &
$5.615$  &  $12.5$  &  $0.0811$  &  $0.935$  &  $8.502$  &  $4.431$  &
$5.8~10^1$\\
8 & $2.0~  (17.02$ km s$^{-1}$) & $15^\circ$ &  $63.2^\circ$ & $0.5$ &
$1.337$ & $1$ & $0.2716$ & $3.305$ & $7.283$ & $52.00$ & $7.6$\\
9 & $2.0~  (17.02$ km s$^{-1}$) & $10^\circ$ &  $64.0^\circ$ & $0.5$ &
$1.421$ & $1$ & $0.2532$ & $3.305$ & $7.423$ & $37.84$ & $6.05$\\
10 & $3.2~  (26.91$ km s$^{-1}$) & $90^\circ$ &  $90^\circ$ & $0.57$ &
$3.869$ & $3.035$ & $0.0436$ & $2.000$ & $13.61$ & $7.275$ & $4.6~10^1$\\
11 & $3.1~ (25.74$ km s$^{-1}$) & $15^\circ$ & $78.5^\circ$ & $0.67$ &
$5.335$  &  $1.542$  &  $0.0296$  & $3.000$  &  $14.55$  &  $4.231$  &
$2.2~10^1$\\ \hline
\end{tabular}
\end{center}
\end{table*}

\subsection{Steady shocks}

Fig.  \ref{fig7} also  shows the path in the $n-p$  plane for a steady
shock and Fig.  \ref{fig8} the structure of its cooling region. It can
be seen  from Fig. \ref{fig8}b  that the flow is  indeed approximately
isobaric in  the unstable region.   From Fig.  \ref{fig8}d we  can see
that  it  is   unstable  for  $-29.934  <  x  <   -13.742$,  but  that
$-\Im(\omega)/|\sigma_c|  <  5.5$.   The analysis  is  therefore  only
marginally valid even for a shock such as this that only just triggers
a phase change.  Note that the  cooling lengths for this shock and the
magnetic version shown in Fig.   \ref{fig9} are very large because the
path in the $n-p$ plane passes  very close to the equilibrium curve at
$n = 1$ where the cooling time is long.

The effect of pure thermal instability in shocks that drive the gas to
the cold stable  state is not very interesting. Although  the gas must
pass through the unstable region in which the instability can grow, we
will see this does not have a dramatic effect on the overall structure
of cooling region. Since the gas ends up in a stable region in which a
two-phase medium is impossible,  any density inhomogeneities generated
by  the instability  must decay.   Colliding flows  that lead  to such
shocks do produce interesting density structures, but these are due to
other  effects such  as  the  thin-shell instability,  Rayleigh-Taylor
instability     or    self-gravity     (e.g.     \citealt{Koyama:2002,
  Heitsch:2008a, Heitsch:2008b, Fogerty:2016})

Thermal instability can only  produce persistent density variations if
the  gas remains  in  the  unstable region  as  it  cools towards  the
equilibrium state, which cannot happen  behind a shock unless there is
a magnetic field.  Fig.  \ref{fig7} also  shows the path of an oblique
MHD  fast shock,  from which  we  can see  that even  a small  initial
magnetic field can lead  to a final state on the  unstable part of the
equilibrium curve.   This is  because there  is enough  compression to
increase the magnetic  field to the point where  the magnetic pressure
dominates, as  can be seen  from Fig.  \ref{fig9}b.   Fig. \ref{fig9}d
also tells  us that  it is  unstable for  $x <  -37.4$ and  the linear
analysis is reasonably accurate since the growth rate is significantly
larger than  $|\sigma_c|$.  Note that  the growth rate is  positive in
the final state,  unlike the shock with a stable  final state shown in
Fig. \ref{fig8}.

Such steady shock solutions in which the gas reaches the unstable part
of the equilibrium curve will obviously not occur in nature and indeed
simulations (e.g.  \citealt{{Koyama:2002}, Audit:2005}) show that they
are  unstable. The  only possibility  is that  the gas  separates into
stable warm and  cold phases if the  gas pressure is in  the range for
which  these phases  can  coexist  (\citealt{Inoue:2009}).  This  also
happens if  one perturbs gas on  the unstable part of  the equilibrium
curve (e.g.  \citealt{Wareing:2019}).  The most interesting shocks are
therefore those  for which the  density in the  final state is  in the
unstable  region  of  the  equilibrium  curve.   The  largest  density
contrast between the phases occurs when the density in the final state
is near the lower end of the unstable region i.e. $n \simeq 1$.

The  steady shock  solutions  are described  by  four parameters:  the
upstream density,  Mach number, plasma  $\beta$ and the  angle between
the  magnetic  field and  the  shock  normal.   There is  therefore  a
three-fold infinity of solutions that can reach any given point on the
equilibrium curve. Table \ref{table1} gives the properties of a number
of such  solutions, for  which all  of the  MHD shocks  end up  on the
unstable part of the equilibrium curve. Of these, we expect shock 5 to
give the largest  density contrast between the phases  since its final
density is closest to the lower stability limit.

The  table also  shows  the amplification  factor  of the  instability
defined by

\vspace*{-10pt}
\begin{equation}
A = \exp \left\{ {\int{ max \left[ {0, - \Im(\omega)} \right]
      \frac{{\rm d} 
    x}{v_x}} } \right\},
  \label{ampfac}
\end{equation}
\vspace*{-5pt}

\noindent
where the  integral is  from the  shock to  the intersection  with the
equilibrium curve.  This definition excludes the damping in the stable
regions,  which means  that the  amplification factor  for the  purely
hydrodynamic shocks is seriously overestimated.

The cooling region  in both the hydrodynamic (shock 1)  and MHD (shock
2) shock  is very  long because  the track in  the $n-p$  plane passes
close to the equilibrium curve, but Table \ref{table1} shows that this
is exceptional.  Stronger shocks  have shorter cooling regions because
their track  is far  from the  equilibrium curve  and MHD  shocks have
lower   densities  and   hence   longer  cooling   regions  than   the
corresponding hydrodynamic ones.  It is also clear from the difference
between the final  density in the hydrodynamic and MHD  cases, that it
does not require  much of an upstream magnetic field  for the magnetic
pressure to dominate in the final state.

The  fact that  even a  very  small magnetic  field has  such a  large
effect,  means  that purely  hydrodynamic  simulations  are of  little
relevance. This is pretty obvious and  has been pointed out by several
authors  (e.g.   \citealt{Hennebelle:2000,  Heitsch:2009,  Inoue:2008,
  Inoue:2009}).   They all  conclude that  the immediate  formation of
very dense gas  in colliding flows is prevented by  a typical magnetic
field  unless  it  is  very  closely aligned  with  the  flow.   Table
\ref{table1} confirms this: the magnetic  field dominates in the final
state for shocks 2, 4, 5 and  7 even though the initial magnetic field
is  implausibly  small.  It  also  dominates  for the  more  realistic
initial field in  shocks 8 to 11,  even for small values  of the angle
between the field and the shock normal.

\section{Numerical calculations}
\label{section_numcalc}

In order to  find out what happens to steady  solutions that reach the
unstable state,  we carried out time  dependent numerical calculations
with the same  AMR MHD code, MG, as in  \cite{Wareing:2016a}.  To keep
things as  simple as  possible, these  were two  dimensional Cartesian
calculations starting with a steady  shock solution propagating in the
$x$ direction.  This is perturbed by  imposing a periodic shift in the
$x$ position of  the shock and hence the whole  solution.  The initial
solution is then given by

\vspace*{-10pt}
\begin{equation}
{\bf p}(x,y) = {\bf p}_s[x + \sin ( 2 \pi y / y_d)],
\label{init_con}
\end{equation}

\noindent
where ${\bf p}_s(x)$ is the steady  solution and $y_d$ is the width of
the domain  in the  $y$ direction.  The  resulting initial  states are
shown  in  Figs.   \ref{fig10},   \ref{fig11}  and  \ref{fig12}.   The
upstream state was  imposed at the right $x$ boundary  and on the left
the $x$ velocity was fixed at that of the end state with zero gradient
for the other variables.  The $y$ boundaries were periodic.

Fig. \ref{fig10}b  shows the density  for the hydrodynamic shock  1 at
$20$ Myrs.  The instability has generated corrugations in the boundary
between the warm and cold gas  and variations in the cold gas density,
which  are then  advected  towards the  left  boundary.  However,  the
density of  the cold  gas only  varies from  $100$ cm$^{-3}$  to $300$
cm$^{-3}$ and these regions are not  in pressure equilibrium as can be
seen from Fig. \ref{fig10}c.  There is  also warm gas next to cold gas
in the region $y < 4$ and $x < 2$, but again these are not in pressure
equilibrium and  the warm phase  is in  the unstable region  above the
equilibrium curve.   It is  clear that the  density variations  in the
cold gas  will reduce  as the pressure  equilibrates and  the unstable
warm gas must turn into stable cold  gas since the gas pressure is too
high for a stable  warm phase to exist.  This is  just telling us that
such  a  shock   cannot  generate  a  two-phase   medium  in  pressure
equilibrium, which is exactly what we would expect.

The difference between shocks 2 and  5 shown in Figs.  \ref{fig11} and
\ref{fig12}, is that shock 5 has a larger pre-shock magnetic field. As
a result,  the magnetic pressure  dominates for  $x < 60$,  whereas in
shock 2 this  does not happen until $x <  15$. The disturbances caused
by the instability are therefore able  to distort the field in shock 2
to produce the  ripples in the field lines in  Fig.  \ref{fig11}b, but
not in shock 5.

In both cases  the region near the left boundary  consists of cold gas
in pressure  equilibrium with the  warm gas, as  can be seen  from the
fact that the filaments are invisible in the plots of the gas pressure
in  Figs.   \ref{fig11}c and  \ref{fig12}c.   It  is also  in  thermal
equilibrium  and is  therefore a  genuine two-phase  medium.  In  both
cases the mass fraction of the  unstable gas near the left boundary is
less than  10\% i.e.  the gas  has largely separated into  stable warm
and cold phases.  This separation into warm and cold phase for shock 5
can clearly be  seen in the plot  of mass fraction in  the $n-p$ plane
shown in Fig.  \ref{fig13}.  Note that the amount of gas in this final
state does not increase systematically  since there is ouflow from the
left boundary.  In reality, the mass in the final state would increase
with time  irrespective of whether  the shock is externally  driven or
due to a  collision between two streams.  The thickness  of the region
in the  two-phase state should  increase by $0.434$ pc  Myr$^{-1}$ for
shock 2 and $2.29$ pc Myr$^{-1}$ for shock 5.

The steady  shock solutions are not  a bad guide to  what happens: the
total pressure is  close to the ram pressure, as  we would expect from
global  momentum balance;  the mean  density is  $3.83$ compared  to a
steady value of  $4.37$ for shock 2 and $1.32$  compared to $1.24$ for
shock 5.  However, the gas pressures in the numerical calculations are
$\simeq 3200$ in both cases, whereas  we would expect $1818$ for shock
2 and $4332$ for shock 5.  As a consequence, the warm density and cold
densities are  $\simeq 0.5$ and $\simeq  80$ in both cases  instead of
$0.27$  and $19$  for shock  2 and  $1.24$ and  $106$ for  shock 5  as
required  by  the  gas  pressures  in  the  steady  solutions.   These
differences in  the gas  pressure are not  surprising since  the final
state in  the unsteady  case is a  two phase medium  as opposed  to an
unstable single phase medium.  Even though the mean density is roughly
the  same, the  gas pressure  in the  two phase  state depends  on the
fractions  of warm  and cold  gas, which  in turn  depend on  the time
history of the instability.

The two phase medium with high density filamentary structures in Figs.
\ref{fig11} and \ref{fig12}  is similar to that  generated by randomly
perturbing  an  initially  unstable  state  (\citealt{Wareing:2016a}).
Despite being produced in very different ways, both the separation, $5
- 10$  pc, and  width,  $\simeq 0.5$  pc, of  the  filaments are  very
similar.   In neither  case does  the separation  seem related  to the
initial perturbation, but  it is a factor of a  few times smaller than
the thermal wavelength at the low  density end of the unstable region.
We might expect this wavelength  to be favoured since Fig. \ref{fig5}a
shows that this is the largest wavelength for which the growth rate is
close to  its maximum.   We have not  included thermal  conduction, so
there  is no  physical  maximally unstable  wavelength, although  very
short wavelengths are suppressed by numerical thermal conductivity.

Without self-gravity, this  two-phase medium would not  evolve as long
as the total pressure remains  constant. Self-gravity is not important
on the  scale of the filaments:  the Jeans length in  the filaments is
$\simeq 12$ pc in both cases,  which is much larger than their widths.
However,  there  is  the  possibility  of  large  scale  gravitational
collapse  along the  field as  in \cite{Wareing:2016a}.   The relevant
timescale for this is the free-fall time for one dimensional collapse,
$1/\sqrt{\pi G \rho}$, which gives $25$ Myr for shock 2 and $42.6$ Myr
for  shock  5.  \cite{Wareing:2016a}  showed  that  most of  the  mass
collects in  a corrugated sheet  perpendicular to the  magnetic field,
which can then collapse perpendicular to the field if the mass to flux
ratio is large enough.

\cite{Fogerty:2016} point  out that  the appropriate critical  mass to
flux ratio is  the one for a  field perpendicular to a  plane layer in
hydrostatic equilibrium

\begin{equation}
\frac{\Sigma}{B} = \frac{1}{2 \pi \sqrt{G}} = 616.25
\mbox{~c.g.s.},
\label{critmass2flux}
\end{equation}

\noindent
where    $\Sigma$   is    the   surface    density   of    the   layer
(\citealt{Nakano:1978}). The two-phase region will be supercritical if
its width along the field is greater than 

\vspace*{-10pt}
\begin{equation}
W_c = 100 \left(\frac{B/10^6}{n} \right) \mbox{~~pc}.
\label{critwidth}
\end{equation}

\noindent
This gives $26$ pc for shock 1 and  $76$ pc for shock 5, so we clearly
need long-lived, large scale flows for gravitational collapse.

This  all  assumes  that  the two-phase  region  is  constrained  from
expanding  perpendicular  to  the  inflow.  This  is  true  for  those
simulations that impose periodic  conditions at the boundaries without
inflow:  \citealt{Koyama:2002, Heitsch:2009,  Inoue:2008, Inoue:2009}.
\cite{Audit:2005}  used  free  boundary conditions  for  their  purely
hydrodynamic calculations, as did \cite{Fogerty:2016} who considered a
parallel  field  with $\beta  =  10$.   Since  the initial  shock  was
effectively  hydrodynamic  in  both  cases, it  produced  high  enough
densities  for self-gravity  to  be significant  despite  the lack  of
constraint on the sideways expansion.

\subsection{Slow shocks}

So far we have only considered fast MHD shocks, but if they are due to
a collision between two streams, then  there must also be slow shocks.
For example,  in a plane  symmetric collision between two  streams the
two fast shocks generate velocities  perpendicular to the shock normal
that have opposite  signs.  At the interface these  velocities must be
equal,  which can  only be  accomplished  by a  slow shock  or a  fast
rarefaction.   In a  plane collision  the only  possibility is  a slow
shock.

\cite{Inoue:2009} find clear evidence of  slow shocks in some of their
cases and not  in others. In fact  slow shocks must be  present in all
cases, but  in some of  them the shocks move  so slowly that  they are
hard to resolve.  For example, Fig.  \ref{fig14} shows the region near
the  interface for  a  one dimensional  version of  their  case 1a:  a
collision  with initial  density  $0.67$, velocity  $20$ km  s$^{-1}$,
field $3$ $\mu$G at an angle of $15^{\circ}$ to the flow.  Shock 11 in
Table  \ref{table1}  is  the  steady  fast  shock  generated  by  this
collision if  we ignore the  slow shock.  We can  see that there  is a
slow shock  at $x =  0.3$ in which  the transverse field  and velocity
decrease so  that the transverse  velocity vanishes at  the interface.
Note that the  oscillations are due to the instabilities  in the state
upstream of the slow shock.  Although  the density behind the shock is
$\simeq 1.75~10^3$, the amount of  mass involved is negligible.  Since
this is generally true, these shocks are of little significance, which
is just  as well since  they are very hard  to resolve in  many cases:
this calculation required  $7$ levels of AMR with  a finest resolution
of $4~10^{-3}$ pc.

\subsection{Shock-cloud interactions}

\cite{Van Loo:2010} considered  a plane fast shock  interacting with a
spherical warm cloud with density  $0.45$ in pressure equilibrium with
a hot medium with density $n = 0.01$.  The shock sonic Mach number was
$2.5$ ($\equiv 142.5$ km s$^{-1}$), the  cloud radius was $200$ pc and
the initial  magnetic field was uniform  with $\beta = 1$.   They used
the  heating and  cooling  prescription described  \cite{Sanchez:2002}
which differs somewhat from the one in \cite{Koyama:2002} that we have
considered here. For  example, it is unstable for $n  \ge 0.5$, rather
$n =  1$. However,  we do  not expect  this to  lead to  a qualitative
difference in the results.

They found that a slow shock  formed at the boundary between the cloud
and the ambient  medium, but it only involved a  significant amount of
mass when  the incident shock  normal was  parallel to the  field: its
effect was  neglible even for  an angle  as small as  $15^\circ$. Slow
shocks can therefore only generate a significant amount of gas at high
densities  when  the field  and  the  shock  normal are  very  closely
aligned.  In  the general  case with plausible  values of  the initial
$\beta$, slow shocks will be unimportant and most of the material will
end up in the two-phase state.

Fig. \ref{fig15} shows  the density and field lines in  a very similar
calculation to  these: the density of  external medium is $n  = 0.01$,
the  thermal Mach  number of  the shock  is $2.5$  ($\equiv 150.4$  km
s$^{-1}$), the cloud radius, $R_c$,  is $200$ pc, the initial pressure
is $3150.25k_B$, the  initial $\beta = 1$ ($B =  3.3~\mu$G), the field
is  parallel to  the $x-y$  plane  at an  angle of  $45^\circ$ to  $x$
axis. The shock travels in the  $-x$ direction.  The domain is $-3 R_c
\le x,  ~y, ~z \le 3  R_c$ and 6 grid  levels were used with  a finest
resolution of $1.25$  pc, which is slightly better than  the $1.67$ pc
in \cite{Van Loo:2010}.   The most significant difference  is that the
energy source function is  given by equations (\ref{energysource}) and
(\ref{cool}).

The regions with  density $\simeq 100$ are curved sheets  about $5$ pc
thick  and an  extent of  about $200$  pc perpendicular  to the  $x-y$
plane. There is  a region between these sheets that  has $\beta < 0.1$
and is in the unstable density range, but above the equilibrium curve,
which  we expect  to cool  and evolve  into a  two-phase medium.   The
sheets have slightly higher gas pressure than this unstable region and
are accumulating mass.   They are on a much larger  scale than that of
thermal  instability  in  Figs.  \ref{fig11}  and  \ref{fig12},  which
suggests that they  are a result of the  large-scale shock propagation
rather than thermal instability.

The Jeans length defined by

\vspace*{-10pt}
\begin{equation}L_J = \left( \frac{\pi a^2}{G \rho} \right)^{1/2},
\label{jeanslength}
\end{equation}

\noindent
is $\simeq 10 - 20$ pc in the sheets, so that self-gravity is becoming
significant. Gravity  is likely to  bring the sheets and  the material
between them  together, which would  make them close  to supercritical
according  to  equation  (\ref{critmass2flux}), especially  since  the
field is mostly  not perpendicular to the sheets. They  should then be
subject  to the  gravitational instabilities  considered by  \cite{Van
  Loo:2014}.

One might  have hoped that  slow shocks would produce  high densities,
but there  is no evidence that  they play a significant  role. This is
consistent  with the  results in  \cite{Van Loo:2010}  for this  angle
between the field and the shock normal.

Although  this  calculation  and  those  in  \cite{Van  Loo:2010}  are
interesting,  there are  two  reasons  why they  must  be regarded  as
indicative rather  accurate solutions  to the  problem as  posed.  The
first is that  the resolution is not sufficient to  resolve the scales
on which  the thermal  instability appears  in Figs.   \ref{fig11} and
\ref{fig12}. The second is that the  flow behind the incident shock is
sub-fast, which means that the  reflected shock propagates to upstream
infinity.   In our  calculation it  reaches the  upstream boundary  at
$7.2$ Myrs,  after which the  external flow is incorrect.   This might
not  actually matter  very much  since  the dynamic  pressures in  the
external flow  are too small to  have much effect on  the evolution of
the cloud.   We tested this by  reducing the size of  the domain after
$12.25$ Myrs and found that this made little difference to the flow in
the cloud.

\section{Summary and conclusions}
\label{section_sum}

In  this paper  we  have  reworked the  linear  stability analysis  in
\cite{Field:1969} using  a combination of the  Hermite-Biehler theorem
and    Whitham's   theory    of    wave    hierarchies   in    Section
\ref{section_hypbal}, analysed its implications  for the energy source
in  \cite{Koyama:2002}   in  Section  \ref{Koyama_en}   and  described
appropriate numerical calculations in Section \ref{section_numcalc}.

Most of the results in Section \ref{section_hypbal} are already known,
but our  method simplifies  the calculations  considerably as  well as
establishing a simple relationship between the dispersion relation and
the various  physical processes.  For  example, we were able  to write
down the  dispersion relation for MHD  simply from a knowledge  of the
MHD wavespeeds  and hence show that  that the inclusion of  a magnetic
field has no effect on the  stability.  We also consider the stability
of non-equilibrium states and show  that the standard analysis is only
quantitatively valid for shocks that  are barely capable of triggering
a transition to the cold phase.

Section  \ref{Koyama_en} discusses  the stability  properties of  both
equilibrium  and non-equilibrium  states  for the  widely used  energy
source function  suggested by \cite{Koyama:2002}.  We  also computed a
number of  steady shock  solutions, both with  and without  a magnetic
field. These results  confirm that for most  plausible parameters, the
final state lies  on the unstable part of the  equilibrium curve. This
means that  such shocks  cannot exist,  but it is  clear that  the end
result must be  a two-phase medium consisting of warm  and cold phases
with  the gas  pressures  in  equilibrium. This  is  confirmed by  the
numerical calculations of perturbed  steady shock solutions in Section
\ref {section_numcalc}.  The main point  here is that the steady shock
solutions are useful for analysing numerical calculations, even though
they cannot exist in reality.

We considered a shock-cloud interacton similar to those in \cite{Van
  Loo:2010}. This large scale shock interaction produces dense sheets
whose scale is determined by the size of the cloud rather than that of
the thermal instability.  They are sufficiently dense to collapse
under their own gravity.

Finally,  we have  shown  that  slow shocks  are  unlikely  to play  a
significant role in these kinds of flow. This is a pity since they are
the only way of producing high  densities in the presence of plausible
magnetic fields.

\section*{Acknowledgements}

We are  grateful to an anonymous  referee for helpful comments  on the
original  version.   This  work  was  supported  by  the  Science  and
Technology  Facilities Council  (STFC,  Research Grant  ST/P00041X/1).
The calculations for this paper were performed on the DiRAC 1 Facility
at Leeds jointly funded by STFC,  the Large Facilities Capital Fund of
BIS and  the University of  Leeds and on  other HPC facilities  at the
University of Leeds.  These facilities are hosted  and enabled through
the ARC HPC resources and support  team at the University of Leeds, to
whom we extend our grateful thanks.   The DiRAC Data Centric system at
Durham  University  was  also  used, operated  by  the  Institute  for
Computational  Cosmology on  behalf  of the  STFC  DiRAC HPC  Facility
(www.dirac.ac.uk).   This  equipment  was  funded by  a  BIS  National
E-infrastructure  capital  grant   ST/K00042X/1,  STFC  capital  grant
ST/K00087X/1,   DiRAC  Operations   grant  ST/K003267/1,   and  Durham
University.  DiRAC is part of the National E-Infrastructure.

\bibliographystyle{mn2e}


\label{lastpage}

\end{document}